\documentclass[journal,twoside,web]{ieeecolor}
\usepackage{generic}
\usepackage{graphics}
\usepackage{subfigure}
\usepackage{amsfonts}
\usepackage{mathrsfs}
\usepackage{epsfig}
\usepackage{times}
\usepackage{amsmath}
\usepackage{amssymb}
\usepackage{algorithm}
\usepackage{ulem}
\usepackage{mathabx}
\usepackage{enumerate}
\usepackage{epstopdf}
\newtheorem{theorem}{Theorem}
\newtheorem{corollary}{Corollary}
\newtheorem{definition}{Definition}

\newtheorem{lemma}{Lemma}
\newtheorem{remark}{Remark}
\newtheorem{assumption}{Assumption}
\def\BibTeX{{\rm B\kern-.05em{\sc i\kern-.025em b}\kern-.08em
 T\kern-.1667em\lower.7ex\hbox{E}\kern-.125emX}}
\begin{document}
\title{\bf Differentially Private Distributed Stochastic Optimization with Time-Varying Sample Sizes\thanks{The work was supported by National Key R\&D Program of China under Grant 2018YFA0703800, National Natural Science Foundation of China under Grant 62203045, T2293770. Corresponding author: Ji-Feng Zhang.}}
\author{Jimin Wang, \IEEEmembership{Member,~IEEE,} and Ji-Feng Zhang, \IEEEmembership{Fellow,~IEEE}
\thanks{Jimin Wang is with School of Automation and Electrical Engineering, University of Science and Technology Beijing, Beijing, 100083, China; (e-mails: jimwang@ustb.edu.cn)}
\thanks{Ji-Feng Zhang is with Key Laboratory of Systems and Control, Institute of Systems Science, Academy of Mathematics and Systems Science, Chinese Academy of Sciences, Beijing 100190, and also with the School of Mathematical Sciences, University of Chinese Academy of Sciences, Beijing 100049, China. (e-mails: jif@iss.ac.cn)}}
\maketitle
\begin{abstract}
Differentially private distributed stochastic optimization has become a hot topic due to the urgent need of privacy protection in distributed stochastic optimization. In this paper, two-time scale stochastic approximation-type algorithms for differentially private distributed stochastic optimization with time-varying sample sizes are proposed using gradient- and output-perturbation methods. For both gradient- and output-perturbation cases, the convergence of the algorithm and differential privacy with a finite cumulative privacy budget $\varepsilon$ for an infinite number of iterations are simultaneously established, which is substantially different from the existing works. By a time-varying sample sizes method, the privacy level is enhanced, and differential privacy with a finite cumulative privacy budget $\varepsilon$ for an infinite number of iterations is established. By properly choosing a Lyapunov function, the algorithm achieves almost-sure and mean-square convergence even when the added privacy noises have an increasing variance. Furthermore, we rigorously provide the mean-square convergence rates of the algorithm and show how the added privacy noise affects the convergence rate of the algorithm. Finally, numerical examples including distributed training on a benchmark machine learning dataset are presented to demonstrate the efficiency and advantages of the algorithms.
\end{abstract}

\IEEEpeerreviewmaketitle

\section{Introduction}\label{sec1}
\IEEEPARstart{I}{n} recent years, information and artificial intelligence technologies are being increasingly employed in emerging applications such as the Internet of Things, cloud-based control systems, smart buildings, and autonomous vehicles \cite{Zhang2021a}. The ubiquitous employment of such technologies provides more ways for an adversary to access sensitive information (e.g., eavesdropping on a communication channel, hacking into an information processing center, or colluding with participants in a system), and thus rapidly increases the risk of privacy leakage. For example, traffic monitoring systems may reveal users' positional trajectories and further disclose details about their driving behavior and frequently visited locations such as the locations of residence and work \cite{Ny2014}. In the electric vehicle market, the leakage of the electric vehicle charging schedule will expose users' living habits and customs, and even violate personal and property safety \cite{Han2017}.  As such, privacy has become a pivotal concern for modern control systems. So far, some privacy-preserving approaches have been recently proposed for control systems relying on homomorphic encryption \cite{Lu2018,Ruan2019}, state decomposition \cite{Wang2019}, and adding artificial noise \cite{Mo2017,He2018,Dwork2006}. Although allowing for computations performed on encrypted data, the communication overhead of homomorphic encryption methods greatly increases with the increase of iterations and agents, which is not practical. Further, the computation results can be revealed only by the private key owner (e.g., an agent or a third party), and thus homomorphic encryption methods typically require a trusted third party \cite{Lu2018,Ruan2019}. Although state decomposition-based methods have small computation loads, they are only suitable for specific systems. Among others, differential privacy is a well-known privacy notion and provides strong privacy guarantees. Thanks to its powerful features, differential privacy has been widely used in deep learning \cite{Abadi2016,Lu2022}, empirical risk minimization \cite{Bassily2014}, stochastic optimization \cite{Song2013}-\cite{Zhang2021b}, distributed consensus \cite{Nozari2017,Liu2020,WangJM2022}, and distributed optimization and game \cite{Ding2021a,Ye2021,Wang2022}.

Distributed (stochastic) optimization has been widely used in various fields, such as  big data analytics, finance, and distributed learning \cite{Nedic2009}-\cite{Reisizadeh2022}. At present, there are many important techniques to solve distributed stochastic optimization, such as stochastic approximation \cite{Doan2021a}-\cite{Reisizadeh2022} and time-varying sample-size. As a standard variance reduction technique, time-varying sample-size schemes have gained increasing research interests and have been widely used to solve various problems, such as large-scale machine learning \cite{Byrd2012}, stochastic optimization \cite{Lei2018}-\cite{Jalilzadeh2022}, and stochastic generalized equations \cite{Cui2023}. In the class of time-varying sample-size schemes, the true gradient is estimated by the average of an increasing number of sampled gradients, which can progressively reduce the variance of the sample-averaged gradients. In distributed stochastic optimization, sensitive personal information is frequently embedded in each agent's sampled gradient. The main reason is that the sampled gradient contains agent-specific data as input, and such data are often private in nature. For example, in smart grid applications, the power consumption data, contained in the sampled gradient, of each household should be protected from being revealed to others because it can demonstrate information regarding the householders (e.g., their activities and even their health conditions such as whether they are disabled or not). In machine learning applications, sampled gradients are directly calculated from and embed the information of sensitive training data. Hence, information regarding the sampled gradient is considered to be sensitive and should be protected from being revealed in the process of solving the distributed stochastic optimization problem.

Privacy-preserving distributed (stochastic) optimization method has recently been studied, including the inherent privacy protection method \cite{Wang2022a}, quantization-enabled privacy protection method \cite{Wang2022d}, and differential privacy method \cite{Wang2022b}-\cite{Gratton2021}. An important result that the convergence and differential privacy with a finite cumulative privacy budget $\varepsilon$ for an infinite number of iterations hold simultaneously has been given for distributed optimization in \cite{Wang2022b}, but this can not be directly used for distributed stochastic optimization. Based on the gradient-perturbation mechanism \cite{Wang2022a} or a stochastic ternary quantization scheme \cite{Wang2022d}, the privacy protection distributed stochastic optimization algorithm with only one iteration was proposed, respectively. Two common methods have been proposed for differential privacy distributed stochastic optimization, namely, gradient-perturbation \cite{Li2018}-\cite{Liu2022b} and output-perturbation \cite{Li2018,Huang2019,Gratton2021}. However, the existing method induces a tradeoff between privacy and accuracy. For the gradient-perturbation case, the mean square convergence of the proposed algorithm cannot be guaranteed, although a finite cumulative privacy budget $\varepsilon$ for an infinite number of iterations has been presented in \cite{Xu2021,Ding2021b,Liu2022b}. For the output-perturbation case, to guarantee the accuracy of the algorithm, $\varepsilon$-differential privacy was proven only for one iteration, leading to the cumulative privacy loss of $k\varepsilon$ after $k$ iterations \cite{Li2018,Huang2019,Gratton2021}. To the best of our knowledge, the convergence of the algorithm and differential privacy with a finite cumulative privacy budget $\varepsilon$ for an infinite number of iterations has not been simultaneously established for distributed stochastic optimization. This observation naturally motivates the following interesting questions. (1) How to design the differentially private distributed stochastic optimization algorithm such that the algorithm protects each agent's sensitive information with a finite cumulative privacy budget $\varepsilon$ and simultaneously guarantees convergence? (2) How does the added privacy noise affect the convergence rate of the algorithm? The current paper mainly aims to answer these two questions.

Two differentially private distributed stochastic optimization algorithms with time-varying sample sizes are proposed in this paper. Both the gradient- and output-perturbation methods are given. The main contributions of this paper are summarized as follows:
\begin{itemize}
\item A differentially private distributed stochastic optimization algorithm with time-varying sample sizes is presented for both output- and gradient-perturbation cases. By a time-varying sample sizes method, the convergence of the algorithm and differential privacy with a finite cumulative privacy budget $\varepsilon$ for an infinite number of iterations can be simultaneously established even when the added privacy noises have an increasing variance. Compared with \cite{Li2018,Xu2021,Ding2021b}, the mean-square and almost sure convergence of the algorithm can be guaranteed for both gradient- and output-perturbation methods. Compared with \cite{Wang2022d}, \cite{Li2018}-\cite{Gratton2021}, a finite cumulative privacy budget $\varepsilon$ for an infinite number of iterations is proven for both gradient- and output-perturbation methods.
\item The mean-square convergence rate of the algorithm with a two-time scale stochastic approximation-type step size is provided by properly selecting a Lyapunov function. Compared with the existing distributed stochastic optimization algorithms with or without privacy protection \cite{Olshevsky2019, Wang2022a,Wang2022d}, we present the mean-square convergence rate of the algorithm. Furthermore, compared with \cite{Doan2021a,Reisizadeh2022}, we give the convergence rate with more general noises.
\end{itemize}
The remaining sections of this paper are organized as follows: Section II introduces the problem formulation. In Sections III and IV, the privacy and convergence analyses for differentially private distributed stochastic optimization with time-varying sample sizes are presented for both output- and gradient-perturbation cases. Section V provides examples on distributed parameter estimation problems, and distributed training of a convolutional neural network over ``MNIST" datasets. Some concluding remarks are presented in Section VI.

{\bf Notations}: Some standard notations are used throughout this paper. $X \geq 0$ ($X>0$) means that the symmetric matrix $X$ is semi-positive definite (positive definite). $\mathbf{1}$ stands for the appropriate-dimensional column vector with all its elements equal one. $\mathbb{R}^{n}$ and $\mathbb{R}^{m\times n}$ denote the $n$-dimensional Euclidean space and the set of all $m\times n$ real matrices, respectively. For any $w$, $v\in\mathbb{R}^{n}$, $\langle w, v\rangle$ denotes the standard inner product on $\mathbb{R}^{n}$. $\|x\|$ refers to the Euclidean norm of vector $x$. $I$, $0$ are an identity matrix and a zero matrix with appropriate dimensions, respectively. For a differentiable function $f(\cdot)$, $\nabla f(w)$ denotes the gradient of $f(\cdot)$ at $w$. The expectation of a random variable $X$ is represented by $\mathbb{E}[X]$. Given two real-valued functions $f(k)$ and $g(k)$ defined on $\mathbb{N}$ with $g(k)$ being strictly positive for sufficiently large $k$, denote $f(k)=O(g(k))$ if there exist $M>0$ and $k_0 > 0$ such that $|f(k)|\leq Mg(k)$ for any $k\geq k_0$; $f(k) = o(g(k))$ if for any $\epsilon >0$ there exists $k_0$ such that $|f(k)|\leq \epsilon g(k)$ for any $k>k_0$. $\lceil x\rceil$ denotes the smallest integer greater than $x$ for $x\in \mathbb{R}$.

\section{Problem formulation}
\subsection{Distributed stochastic optimization}
Consider the following optimization problems defined over a network, which needs to be distributedly solved by $n$ agents:
\begin{eqnarray}\label{problem}
\left.
  \begin{array}{ccc}
     \underset{x\in \mathbb{R}^{d}}{\min}  & f(x)=\sum_{i=1}^{n}f_{i}(x), &f_{i}(x)\triangleq\mathbb{E}_{\xi_{i}\sim\mathcal{D}_{i}}[\ell_{i}(x,\xi_{i})].\\
  \end{array}
\right.
\end{eqnarray}
where $x$ is common for any $i\in\mathcal{V}$, but $\ell_{i}$ is a local cost function private to Agent $i$, and $\xi_{i}$ is a random variable. $\mathcal{D}_{i}$ is the local distribution of the random variable $\xi_{i}$, which usually denotes a data sample in machine learning. The following assumptions are presented to ensure the well-posedness of (\ref{problem}):
\begin{assumption}\label{assum1} For any $i\in\mathcal{V}$, each function $\nabla f_{i}$ is Lipschitz continuous, i.e., there exists $L_{i}>0$ such that
\begin{equation*}
\|\nabla f_{i}(x)-\nabla f_{i}(y)\| \leq L_{i}\|x-y\|, \forall x, y \in \mathbb{R}^{d}.
\end{equation*}
each function $f_{i}$ is $\mu$-strongly convex if and only if $f_{i}$ satisfies
\begin{equation*}
\langle \nabla f_{i}(x)-\nabla f_{i}(y), x-y\rangle\geq \mu\|x-y\|^{2}, \forall x, y \in \mathbb{R}^{d}.
\end{equation*}
\end{assumption}
\subsection{Distributed subgradient methods}
Distributed subgradient methods for solving the distributed (stochastic) optimization problem were first studied and rigorously analyzed by \cite{Nedic2009,Nedic2010}. In these algorithms, each agent $i$ iteratively updates its decision variables $x_{i}$ by combining an average of the states of its neighbors with a gradient step as follows:
$x_{i,k+1}=\sum_{j\in\mathcal{N}_{i}}a_{ij}x_{j,k}-\alpha_{k}g_{i}(x_{i,k})$, where $\alpha_{k}$ is the time-varying step size corresponding to the influence of the gradients on the state update rule at each time step. Considering the randomness in $\ell_{i}(x,\xi_{i})$, the gradient $g_{i}(x_{i,k})$ that can be obtained by each agent $i$ is subject to noises. To reduce the variance of the gradient observation noise, the time-varying sample sizes are used in \cite{Lei2020}. In this case, the gradient that Agent $i$ has for optimization at iteration $k$ is denoted as $\frac{1}{\gamma_{k}}\sum_{l=1}^{\gamma_{k}}g_{i}(x_{i,k},\xi_{i}^{l})$, and $\gamma_{k}>1$ is the number of the sampling gradients used at time $k$, and $\xi_{i}^{l}, l=1,\cdots,\gamma_{k}$ represents the realizations of $\xi_{i}$. For the sake of notational simplicity, $\frac{1}{\gamma_{k}}\sum_{l=1}^{\gamma_{k}}g_{i}(x_{i,k},\xi_{i}^{l})$ is abbreviated as $g_{i}^{k}$. In this paper, the following standard assumption was made about $g_{i}(x_{i,k},\xi_{i}^{l})$:
\begin{assumption}\label{assum2}
For any fixed $l$ and $x_{i,k}\in \mathbb{R}^{d}$, there exists a positive constant $\sigma_{g}$  such that $g_{i}(x_{i,k},\xi_{i}^{l})$ satisfies $\mathbb{E}[g_{i}(x_{i,k},\xi_{i}^{l})]=\nabla f_{i}(x_{i,k})$ and $\mathbb{E}[\|g_{i}(x_{i,k},\xi_{i}^{l})-\nabla f_{i}(x_{i,k})\|^{2}]\leq\sigma_{g}^{2}.$
\end{assumption}

The communication topology $\mathcal{G}=(\mathcal{V},\mathcal{E})$ consists of a non-empty agent set $\mathcal{V}=\{1,2,\cdots,n\}$ and an edge set $\mathcal{E}\subseteq\mathcal{V} \times \mathcal{V}$. $A=[a_{ij}]$ is the adjacency matrix of $\mathcal{G}$, where $a_{ii}>0$ and $a_{ij}>0$ if $(i,j)\in\mathcal{E}$ and $a_{ij}=0$, otherwise. $\mathcal{N}_{i}=\{j\in\mathcal{V}, (j,i)\in\mathcal{E}\}$ denotes the neighborhood of Agent $i$ including itself. $\mathcal{G}$ is called connected if for any pair agents $(i_1, i_{l})$, there exists a path from $i_1$ to $i_{l}$ consisting of edges $(i_1, i_2), (i_2, i_{3}), \cdots, (i_{l-1}, i_{l})$.
\begin{assumption}\label{assum2a} The undirected communication topology $\mathcal{G}$ is connected, and the adjacency matrix $A$ satisfies the following conditions:
(i) There exists a positive constant $\eta$ such that $a_{ij}>\eta$ for $j\in\mathcal{N}_{i}$, $a_{ij}=0$ for $j\not\in\mathcal{N}_{i}$; (ii) $A$ is doubly stochastic, namely, ${\bf 1}^{T}A={\bf 1}^{T}$, $A{\bf 1}={\bf 1}$.
\end{assumption}

It is considered that the following passive attackers exist in distributed stochastic optimization that have been widely used in the existing works \cite{Wang2022,Wang2022a,Wang2022d}:

\begin{itemize}
\item {\it Semi-honest agents} are assumed to follow the specified protocol and perform the correct computations. However, they may collect all intermediate and input/output information in an attempt to learn sensitive information about the other agents.

\item {\it External eavesdroppers} are adversaries who steal information through wiretapping all communication channels and intercepting exchanged information between agents.
\end{itemize}

Due to the information exchange in the above-mentioned algorithm, the potential passive attackers can always collect $x_{i,k}$ at each time $k$. Meanwhile, the attackers know the topology graph ($A$) and step-size ($\alpha_{k}$). Combining all the information, it is easy for the potential passive attackers to infer the agents' sampled gradients. In this case, raw data directly computes the sampled gradients, further leaking the agents' sensitive information. Therefore, in this paper, privacy is defined as preventing agents' sampled gradients from being inferable by potential passive attackers.
\subsection{Differential privacy}
This subsection presents some preliminaries of differential privacy. In distributed
stochastic optimization algorithms, preserving differential privacy is equivalent to hiding changes in the samples of the gradient information. Changes in the samples of the gradient information can be formally defined by a symmetric binary relation between two datasets called the adjacency relation. Inspired by \cite{Bassily2019,Wang2022}, the following definition is given.
\begin{definition}\label{def1} (Adjacent relation): Given a positive constant $C$, two different samples of the gradients $D_{k}=\{g_{i}(x_{i,k},\xi_{i}^{l}), l=1,2,\cdots,\}$, $D'_{k}=\{g_{i}(x_{i,k},\xi_{i}^{l^{'}}), l^{'}=1,2,\cdots,\}$ are said to be adjacent if they differ in exactly one data sample $l_{0},l^{'}_{0}$ such that $\|g_{i}(x_{i,k},\xi_{i}^{l_{0}})-g_{i}(x_{i,k}, \xi_{i}^{l'_{0}})\|_{1}\leq C$.
\end{definition}
\begin{remark}
Adjacent relation indicates the specific sensitive information that needs to be protected in this paper. From Definition \ref{def1} it follows that $D_{k}$ and $D'_{k}$ are adjacent if only one data sample $l_{0},l^{'}_{0}$ satisfies $\|g_{i}(x_{i,k},\xi_{i}^{l_{0}})-g_{i}(x_{i,k}, \xi_{i}^{l'_{0}})\|_{1}\leq C$ and the others satisfy $\|g_{i}(x_{i,k},\xi_{i}^{l})-g_{i}(x_{i,k}, \xi_{i}^{l'})\|_{1}=0$.
\end{remark}
\begin{definition}\label{def2} \cite{Ny2014} (Differential privacy). Given $\varepsilon\geq0$, a randomized algorithm $\mathcal{A}$ is $\varepsilon$-differentially private at $k$th iteration if for all adjacent $D_{k}$ and $D'_{k}$, and for any subsets of outputs $\Upsilon \subseteq {\rm Range}(\mathcal{A})$ such that
$\mathbb{P}\{\mathcal{A}(D_{k})\in\Upsilon\}\leq  e^{\varepsilon}\mathbb{P}\{\mathcal{A}(D'_{k})\in\Upsilon\}.$
\end{definition}
\begin{remark} The basic idea of differential privacy is to ``perturb" the exact result before release. In this case, an adversary cannot tell from the output of $D_{k}$ with a high probability that an agent's sensitive information has changed or not. The constant $\varepsilon$ measures the privacy level of the randomized algorithm $\mathcal{A}$, i.e., a smaller $\varepsilon$ implies a better privacy level.
\end{remark}

{\bf Problem of interest}: This paper mainly seeks to develop two privacy-preserving distributed stochastic optimization algorithms such that each agent's sensitive information can be protected to a greater extent, and the convergence of the algorithm is guaranteed simultaneously.

\section{Differentially private distributed stochastic optimization via output-perturbation}
In this subsection, a differentially private distributed stochastic optimization algorithm with time-varying sample sizes is presented via output perturbation. Specifically, in each iteration of Algorithm \ref{al1}, rather than its original state, each agent $i$ sends its current noisy state $x_{i,k}+n_{i,k}$ to each of its neighbors $j\in\mathcal{N}_{i}$, where $x_{i,k}$ is the estimate state of Agent $i$ at time $k$, $n_{i,k}\in\mathbb{R}^{d}$ is temporally and spatially independent, and each element is the zero-mean Laplace noise with the variance of $2\sigma_{k}^2$. 
\begin{algorithm}
{\bf Initialization:} Set $k=0$, $x_{i,0}\in\mathbb{R}^{d}$ is any arbitrary initial value for any $i\in\mathcal{V}$.\\
{\bf Iterate until convergence}. Each agent $i\in\mathcal{V}$ updates its state as follows:
\begin{eqnarray}\label{PP1}
x_{i,k+1}
=(1-\beta_{k})x_{i,k}+\beta_{k}\sum_{j\in\mathcal{N}_{i}}a_{ij}(x _{j,k}+n_{j,k})-\alpha_{k}g_{i}^{k},
\end{eqnarray}
where $\alpha_{k}>0$ is the step-size for the gradient step, a new step-size $0<\beta_{k}<1$ is introduced that determines the degree to which information from the neighbors should be weighed, and $n_{j,k}$ is the added privacy noises for Agent $j$ at each time $k$.
\caption{Differentially private distributed stochastic optimization with time-varying sample sizes via output perturbation \label{al1}}
\end{algorithm}
\subsection{Privacy analysis}
This subsection demonstrates the $\varepsilon$-differential privacy of Algorithm 1. We first derive conditions on the noise variances under which Algorithm 1 satisfies $\varepsilon$-differential privacy for an infinite number of iterations. A critical quantity determines how much noise should be added to each iteration for achieving $\varepsilon$-differential privacy, referred to as sensitivity.

\begin{definition}\label{def3} \cite{Han2017} (Sensitivity). The sensitivity of an output map $q$ at $k$th iteration is defined as
$$\Delta_{k}=\sup \limits_{D_{k},D'_{k}:{\rm Adj}(D_{k},D'_{k})}\|q(D_{k})-q(D'_{k})\|_{1}.$$
\end{definition}
\begin{remark}
The sensitivity of an output map $q$ means that a single sampling gradient can change the magnitude of the output map $q$. It should be pointed out that $q$ refers to $x_{i,k}$ for Algorithm 1, and $g_{i}^{k}$ for Algorithm 2.
\end{remark}
\begin{lemma}\label{lem2}
The sensitivity of Algorithm 1 at $k$th iteration satisfies
\begin{equation}\label{sensitivitylemma}
\Delta_{k}\leq\left\{
\begin{array}{lcc}
 \frac{C\alpha_{0}}{\gamma_{0}},&  &k=1;   \\
\sum_{l=0}^{k-2}\prod_{t=l+1}^{k-1}(1-\beta_{t})\frac{C\alpha_{l}}{\gamma_{l}},& & k>1.   \\
 \end{array}
 \right.
\end{equation}
\end{lemma}
\noindent
\textit{Proof}: Recall in Definition \ref{def1}, $D_{k}$ and $D'_{k}$ are any two different samples of the gradient information differing in one data sample at $k$th iteration. $x_{i,k}$ is computed based on $D_{k}$, while $x'_{i,k}$ is calculated based on $D'_{k}$.
For $D_{k}$ and $D'_{k}$, we have
\begin{align}\label{8}
&\|x_{i,k}-x'_{i,k}\|_{1}\cr
\leq&\|(1-\beta_{k-1})(x_{i,k-1}-x'_{i,k-1})\cr
&-\frac{\alpha_{k-1}}{\gamma_{k-1}}(g_{i}(x_{i,k-1},\xi_{i}^{l_{0}})-g_{i}(x_{i,k-1},\xi_{i}^{l'_{0}}))\|_{1}
\cr
\leq&\|(1-\beta_{k-1})(x_{i,k-1}-x'_{i,k-1})\|_{1}+\frac{C\alpha_{k-1}}{\gamma_{k-1}}.
\end{align}
From (\ref{8}) it follows that $\|x_{i,k}-x'_{i,k}\|_{1}=\frac{C\alpha_{0}}{\gamma_{0}},$ when $k=1$; when $k>1$, $\|x_{i,k}-x'_{i,k}\|_{1}=\sum_{l=0}^{k-2}\prod_{t=l+1}^{k-1}(1-\beta_{t})
\frac{C\alpha_{l}}{\gamma_{l}}.$  $\hfill\Box$
\begin{remark} Motivated by \cite{Li2018}, the time-varying sample-size method is used to process multiple samples at the same iteration. Most importantly, the time-varying sample-size method has a great advantage in guaranteeing differential privacy for Algorithm 1. Observing the proof of Lemma \ref{lem2}, it is found that parameter $\frac{1}{\gamma_{k}}$ has reduced the sensitivity of Algorithm 1 and further enhances the privacy protection ability.
\end{remark}
\begin{theorem}\label{thm1} Let $C$ be any given positive number. If $\varepsilon=\sum_{k=1}^{\infty}\frac{\Delta_{k}}{\sigma_{k}},$
then Algorithm 1 is $\varepsilon$-differentially private for an infinite number of iterations.
\end{theorem}
\textit{Proof}: The proof is similar to Theorem 3.5 in \cite{Liu2020}, and thus is omitted here. $\hfill\Box$
\begin{theorem}\label{thm2} Let $\alpha_{k}=\frac{a_{1}}{(k+a_{2})^{\alpha}}$, $\beta_{k}=\frac{a_{1}}{(k+a_{2})^{\beta}}$, $\gamma_{k}=\lceil a_3(k+a_{2})^{\gamma}\rceil$, and $\sigma_{k}=\underline{b}(k+a_{2})^{\eta}$, $0<\beta\leq1,$ $0<\alpha\leq1$, $\gamma\geq0$, $\eta\geq0$, $0<a_1<a_2^\beta$, $a_2,a_3,\underline{b}>0$. If one of the following conditions holds,
\begin{enumerate}[i)]
\item  $\beta=1,\alpha+\gamma-a_{1}<1, \alpha+\gamma+\eta>2$;
\item  $\beta=1,\alpha+\gamma-a_{1}\geq1,a_{1}+\eta>1$;
\item  $0<\beta<1,\alpha+\gamma-\beta+\eta>1$,
\end{enumerate}
then Algorithm 1 is differentially private with a finite cumulative
privacy budget $\varepsilon$ for an infinite number of iterations.
\end{theorem}
\textit{Proof}: We only need to prove that cumulative privacy budget $\varepsilon$ is finite for all $k>1$.
When $\beta=1$, note that $\alpha_{k}=\frac{a_{1}}{(k+a_{2})^{\alpha}}$, $\beta_{k}=\frac{a_{1}}{k+a_{2}}$, $\gamma_{k}=\lceil a_{3}(k+a_{2})^{\gamma}\rceil$, from (\ref{sensitivitylemma}) it follows that
\begin{eqnarray*}
\Delta_{k}\leq\sum_{l=0}^{k-2}\prod\limits_{t=l+1}^{k-1}(1-\frac{a_{1}}{t+a_{2}})\frac{Ca_{1}}{a_{3}(l+a_{2})^{\alpha+\gamma}},\quad k>1.
\end{eqnarray*}
For $k>1$, from Lemma \ref{lemma:prod_est2} it follows that
\begin{align*}
\Delta_{k}=
\left\{
\begin{array}{lcl}
O\left((k+a_{2})^{-\alpha-\gamma+1}\right),  & \alpha+\gamma-a_{1}<1;  \\
 O\left((k+a_{2})^{-a_{1}}\ln k\right),  & \alpha+\gamma-a_{1}=1;   \\
 O\left((k+a_{2})^{-a_{1}}\right), & \alpha+\gamma-a_{1}>1.   \\
\end{array} \right.
\end{align*}
Furthermore, since $\sigma_{k}=\underline{b}(k+a_{2})^{\eta}$, we have
\begin{align*}
\sum_{k=2}^{\infty}\frac{\Delta_{k}}{\sigma_{k}}
=
\left\{\!\!
\begin{array}{lcl}
O\left(\sum\limits_{k=2}^{\infty}(k+a_{2})^{-\alpha-\gamma-\eta+1}\right),  & \alpha+\gamma-a_{1}<1;  \\
 O\left(\sum\limits_{k=2}^{\infty}(k+a_{2})^{-a_{1}-\eta}\ln k\right),  & \alpha+\gamma-a_{1}=1;   \\
 O\left(\sum\limits_{k=2}^{\infty}(k+a_{2})^{-a_{1}-\eta}\right), & \alpha+\gamma-a_{1}>1.   \\
\end{array}\right.
\end{align*}
From Lemma \ref{lemma:prod_est2}, when $\alpha+\gamma-a_{1}<1, \alpha+\gamma+\eta>2$ or $\alpha+\gamma-a_{1}\geq1,a_{1}+\eta>1$, we have $\varepsilon=O\left(1\right)$.

When $0<\beta<1$, from (\ref{sensitivitylemma}) it follows that
\begin{eqnarray*}
\Delta_{k}\leq\sum_{l=0}^{k-2}\prod_{t=l+1}^{k-1}\left(1-\frac{a_{1}}{(t+a_{2})^\beta}\right)
\frac{Ca_{1}}{a_{3}(l+a_{2})^{\alpha+\gamma}}, \quad k>1.
\end{eqnarray*}
By using Lemma \ref{lemma:prod_est}, we have
\begin{align}\label{Cp14}
\Delta_{k}=&O\left(\sum_{l=0}^{k-2}\exp\left(-\frac{a_{1}}{1-\beta}(k+a_{2})^{1-\beta}\right)\right.\cr
&\!\!\left.\cdot\exp\left(\frac{a_{1}}{1-\beta}(l+a_{2})^{1-\beta}\right)\frac{Ca_{1}}{a_{3}(l+a_{2})^{\alpha+\gamma}}\right).
\end{align}
From (\ref{Cp14}) and Lemma \ref{lemma:calc_sum_exp} it follows that
\begin{align*}
\Delta_{k}=&O\left(\exp\left(-\frac{a_{1}}{1-\beta}(k+a_{2})^{1-\beta}\right)\right.\cr
&\!\!\left.\cdot\frac{1}{(k+a_{2})^{\alpha+\gamma-\beta}}
\exp\left(\frac{a_{1}}{1-\beta}(k+a_{2})^{1-\beta}\right)\right)\cr
=& O\left((k+a_{2})^{-\alpha-\gamma+\beta}\right).
\end{align*}
Further, from Lemma \ref{lemma:prod_est2} it follows that when $0<\beta<1$, we have
\begin{align*}
\sum_{k=2}^{\infty}\frac{\Delta_{k}}{\sigma_{k}}
=&O\left(\sum_{k=1}^{\infty}(k+a_{2})^{-\alpha-\gamma+\beta-\eta}\right)\cr
=&\left\{
\begin{array}{lcl}
O\left((k+a_{2})^{-\alpha-\gamma+\beta-\eta+1}\right),  & \alpha+\gamma-\beta+\eta<1;  \\
 O\left(\ln k\right),  & \alpha+\gamma-\beta+\eta=1;  \\
O\left(1\right), & \alpha+\gamma-\beta+\eta>1.   \\
\end{array}\right.
\end{align*}
Based on the above-mentioned discussion, when $\beta=1,\alpha+\gamma-a_{1}<1, \alpha+\gamma+\eta>2$, $\beta=1,\alpha+\gamma-a_{1}\geq1,a_{1}+\eta>1$, or $0<\beta<1,\alpha+\gamma-\beta+\eta>1$ holds, cumulative privacy budget $\varepsilon$ is finite for an infinite number of iterations. $\hfill\Box$
\begin{remark} Theorem \ref{thm2} gives a guidance for choosing $\alpha$, $\beta$, $\gamma$, and $\eta$ to achieve the differentially private with a finite cumulative privacy budget $\varepsilon$ for an infinite number of iterations of Algorithm 1. $\varepsilon$-differential privacy was proven only for one iteration in \cite{Wang2022d,Li2018,Huang2019}, leading to the cumulative privacy loss of $k\varepsilon$ after $k$ iterations, and hence the cumulative privacy budget growing to infinity with time. Therefore, $\varepsilon$ for an infinite number of iterations is smaller in this paper than the ones in \cite{Wang2022d,Li2018,Huang2019}. This implies that the algorithm achieves a better level of privacy protection than the ones in \cite{Wang2022d,Li2018,Huang2019}.
\end{remark}

\subsection{Convergence analysis}
To facilitate convergence analysis of Algorithm 1, the stacked vectors are defined as follows:
$x_{k}=[x_{1,k}, \cdots, x_{n,k}]^{T}, n_{k}=[n_{1,k}, \cdots, n_{n,k}]^{T}, G(x_{k})=[(g_{1}^{k}), \cdots, (g_{n}^{k})]^{T}.$
Let $\overline{x}_{k},\overline{n}_{k} \in\mathbb{R}^{d}$ be the average of $x_{i,k}$, $n_{i,k},$ respectively, i.e., $\overline{x}_{k}=\frac{1}{n}\sum_{i=1}^{n}x_{i,k}=\frac{1}{n}x^{T}_{k}{\bf 1}$,
$\overline{n}_{k}=\frac{1}{n}\sum_{i=1}^{n}n_{i,k}$. Additionally, we use the following notation $W=I-\frac{1}{n}{\bf 1}{\bf 1}^{T}$, $U_{k}=\overline{x}_{k}-x^{*}$, $Y_{k}=x_{k}-{\bf 1}\overline{x}_{k}^{T}=Wx_{k}$. Define $\sigma$-algebra $\mathcal{F}_{k}=\sigma\{G(x_{t}),n_{t},0\leq t\leq k-1\}$.
Then, the compact form of (\ref{PP1}) can be rewritten as follows:
\begin{eqnarray}\label{CA2}
x_{k+1}=(1-\beta_{k})x_{k}+\beta_{k}A(x_{k}+n_{k})-\alpha_{k}G(x_{k}).
\end{eqnarray}
Since $A$ is doubly stochastic, we have
\begin{eqnarray}\label{CA3}
\overline{x}_{k+1}=(1-\beta_{k})\overline{x}_{k}+\beta_{k}(\overline{x}_{k}+\overline{n}_{k})-\frac{\alpha_{k}}{n}\sum_{i=1}^{n}g_{i}^{k}.
\end{eqnarray}

Before discussing the convergence property of the algorithm, the following assumption is presented.
\begin{assumption}\label{assum3}
The step sizes $\alpha_{k}, \beta_{k}$, privacy noise parameters $\sigma_{k}$, and time-varying sample sizes $\gamma_{k}$ satisfy the following conditions:
\begin{itemize}
\item
$\sup_{k}\frac{\alpha_{k}}{\beta_{k}}\leq\min\{\frac{2(1-\sigma_{2})}{3\mu},\frac{(1-\sigma_{2})^{2}\mu^{2}\beta_{0}}{16(6L^{2}\alpha_{0}+n(1-\sigma_{2})\mu\beta_{0})(\beta_{0}+1)L^{2}}\},\\
\sum_{k=0}^{\infty}\frac{\alpha^{2}_{k}}{\beta_{k}}<\infty,
\sum_{k=0}^{\infty}\beta_{k}^{2}\sigma_{k}^{2}<\infty,\\
\sum_{k=0}^{\infty}\frac{\alpha^{2}_{k}}{\gamma_{k}\beta_{k}}<\infty,
\sum_{k=0}^{\infty}\frac{\alpha^{2}_{k}}{\gamma_{k}}<\infty.$
\end{itemize}
\end{assumption}
\begin{remark}  Assumption~\ref{assum3} is satisfied for many kinds of step-sizes and noise parameters. For example, for sufficiently large $a_{2}$, Assumption \ref{assum3} is satisfied in the form of $\alpha_{k}=(k+a_{2})^{-1}$, $\beta_{k}=(k+a_{2})^{-\beta}$, $\beta\in(1/2,1)$, $\sigma_{k}=(k+a_{2})^{\eta}, \eta<\beta-1/2$, $\gamma_{k}=\lceil(k+a_{2})^{\gamma}\rceil, \gamma\geq0$.  Especially, when $\sigma_{k}$ and $\gamma_{k}$ are constants, Assumption~\ref{assum3} becomes the commonly used two-time scale stochastic approximation step-size \cite{Doan2021a,Doan2021b}.
\end{remark}
Next, we provide the mean-square and almost sure convergence of Algorithm 1.
\begin{theorem}\label{alth3} If Assumptions \ref{assum1}-\ref{assum3} hold, then Algorithm 1 converges in mean-square and almost surely for any $i\in \mathcal{V}$, i.e., there exists an optimal solution $x^{*}$ such that $\lim_{k\rightarrow\infty} \mathbb{E}[\|x_{i,k}-x^{*}\|^{2}]=0,$ and $\lim_{k\rightarrow\infty} x_{i,k}=x^{*},$ a.s. $\forall i\in \mathcal{V}.$
\end{theorem}
\noindent
\textit{Proof}: There are three steps for completing the proof. First, the relationships for $\mathbb{E}\left[\|x_{k}-{\bf 1}\overline{x}_{k}^{T}\|^{2}|\mathcal{F}_{k}\right]$ and $\mathbb{E}\left[\|\overline{x}_{k}-x^{*}\|^{2}|\mathcal{F}_{k}\right]$ are, respectively, established in {\it Step 1} and {\it Step 2} as follows:

{\it Step 1:} Note that $WA=AW$ by Assumption \ref{assum2a}. Then, from (\ref{CA2}) and (\ref{CA3}) it follows that
\begin{align}\label{CA4}
Y_{k+1}
=&Wx_{k+1}\cr
=&(1-\beta_{k})Y_{k}+\beta_{k}AW(x_{k}+n_{k})-\alpha_{k}WG(x_{k})\cr
=&(1-\beta_{k})Y_{k}+\beta_{k}AY_{k}+\beta_{k}AWn_{k}-\alpha_{k}WG(x_{k}).
\end{align}
Note that the second largest singular value of $A$ is less than 1 by Assumption \ref{assum2a} (i.e. $0<\sigma_{2}<1$). Then, the following Cauchy-Schwarz inequality holds for some $\eta=(1-\sigma_{2})\beta_{k}>0$ and $a,b\in\mathbb{R}$: $(a+b)^{2}\leq(1+\eta)a^{2}+(1+\frac{1}{\eta})b^{2}.$
Then, by taking the 2-norm square of (\ref{CA4}) and using Cauchy-Schwarz inequality, we have
\begin{align}\label{CA6}
&\|Y_{k+1}\|^{2}\cr
\leq&(1+(1-\sigma_{2})\beta_{k})\|(1-\beta_{k})Y_{k}+\beta_{k}AY_{k}+\beta_{k}AWn_{k}\|^{2}\cr
&+(1+\frac{1}{(1-\sigma_{2})\beta_{k}})\|\alpha_{k}WG(x_{k})\|^{2}\cr
\leq&(1+(1-\sigma_{2})\beta_{k})\|(1-\beta_{k})Y_{k}+\beta_{k}AY_{k}+\beta_{k}AWn_{k}\|^{2}\cr
&+(1+\frac{1}{(1-\sigma_{2})\beta_{k}})\|\alpha_{k}G(x_{k})\|^{2},
\end{align}
where the last inequality used the fact $\|W\|=1$. Next, we analyze each term on the right-hand side of the above inequality. Set $\nabla f(x_{k})=\left[
                   \begin{array}{cccc}
                     \nabla f_{1}(x_{1,k}),& \cdots, &\nabla f_{n}(x_{n,k}) \\
                   \end{array}
                 \right]^{T}$. Then, we have
\begin{align}\label{CA7}
\|G(x_{k})\|^{2}=&\|G(x_{k})-\nabla f(x_{k})+\nabla f(x_{k})\|^{2}\cr
\leq&2\|G(x_{k})-\nabla f(x_{k})\|^{2}+2\|\nabla f(x_{k})\|^{2}.
\end{align}
Denote $X^{*}={\bf 1}\otimes x^{*}$ and $L^{2}=\sum_{i=1}^{n}L_{i}^{2}$. Then, adding and sub\-tract\-ing $\nabla f(X^{*})$ to $\nabla f(x_{k})$, from Assumption \ref{assum1} it follows that
\begin{align}\label{CA8}
&\|\nabla f(x_{k})\|^{2}\cr
\leq&2\|\nabla f(x_{k})-\nabla f(X^{*})\|^{2}+2\|\nabla f(X^{*})\|^{2}\cr
\leq&2L^{2}\|x_{k}-X^{*}\|^{2}+2\|\nabla f(X^{*})\|^{2}\cr
\leq&2L^{2}\sum_{i=1}^{n}\|x_{i,k}-\overline{x}_{k}+\overline{x}_{k}-x^{*}\|^{2}+2\|\nabla f(X^{*})\|^{2}\cr
\leq&4L^{2}\sum_{i=1}^{n}\|x_{i,k}-\overline{x}_{k}\|^{2}+4nL^{2}\|\overline{x}_{k}-x^{*}\|^{2}\cr
&+2\|\nabla f(X^{*})\|^{2}.
\end{align}
From Assumption \ref{assum2} it follows that
\begin{eqnarray}
\mathbb{E}[\|G(x_{k})-\nabla f(x_{k})\|^{2}|\mathcal{F}_{k}]
\leq\frac{n\sigma_{g}^{2}}{\gamma_{k}}.
\end{eqnarray}
In addition, by using Lemma \ref{lemCa3}, we have
\begin{eqnarray}\label{CA9}
\|(1-\beta_{k})Y_{k}+\beta_{k}AY_{k}\|^{2}\leq\|(1-(1-\sigma_{2})\beta_{k})Y_{k}\|^{2},
\end{eqnarray}
Recall that $\mathbb{E}[n_{k}|\mathcal{F}_{k}]=0$. Then, taking the condition expectation of (\ref{CA6}) with respect to $\mathcal{F}_{k}$, from (\ref{CA6})-(\ref{CA9}) it follows that
\begin{align}\label{CA10}
&\mathbb{E}[\|Y_{k+1}\|^{2}|\mathcal{F}_{k}]\cr
\leq&(1+(1-\sigma_{2})\beta_{k})(1-(1-\sigma_{2})\beta_{k})^{2}\|Y_{k}\|^{2} \cr
&+(1+(1-\sigma_{2})\beta_{k})\mathbb{E}[\|\beta_{k}AWn_{k}\|^{2}|\mathcal{F}_{k}]\cr
&+(1+\frac{1}{(1-\sigma_{2})\beta_{k}})\mathbb{E}[\|\alpha_{k}G(x_{k})\|^{2}|\mathcal{F}_{k}]\cr
\leq&(1+(1-\sigma_{2})\beta_{k})(1-(1-\sigma_{2})\beta_{k})^{2}\|Y_{k}\|^{2} \cr
&+(1+(1-\sigma_{2})\beta_{k})\mathbb{E}[\|\beta_{k}AWn_{k}\|^{2}|\mathcal{F}_{k}]\cr
&+(1+\frac{1}{(1-\sigma_{2})\beta_{k}})\alpha^{2}_{k}(8L^{2}\sum_{i=1}^{n}\|x_{i,k}-\overline{x}_{k}\|^{2}\cr
&+8nL^{2}\|\overline{x}_{k}-x^{*}\|^{2}+4\|\nabla f(X^{*})\|^{2}+\frac{2n\sigma_{g}^{2}}{\gamma_{k}})\cr
\leq&(1-(1-\sigma_{2})\beta_{k})\|Y_{k}\|^{2}+(1+\beta_{0})\sigma_{2}^{2}\beta_{k}^{2}\mathbb{E}[\|n_{k}\|^{2}|\mathcal{F}_{k}]\cr
&+\frac{\beta_{0}+1}{(1-\sigma_{2})\beta_{k}}\alpha^{2}_{k}(8L^{2}\sum_{i=1}^{n}\|x_{i,k}-\overline{x}_{k}\|^{2}\cr
&+8nL^{2}\|\overline{x}_{k}-x^{*}\|^{2}+4\|\nabla f(X^{*})\|^{2}+\frac{2n\sigma_{g}^{2}}{\gamma_{k}}).
\end{align}
Note that $\mathbb{E}[\|n_{k}\|^{2}|\mathcal{F}_{k}]=2nd\sigma_{k}^{2}$. Then, from (\ref{CA10}) it follows that
\begin{align}\label{CA10a}
&\mathbb{E}[\|Y_{k+1}\|^{2}|\mathcal{F}_{k}]\cr
\leq&\|Y_{k}\|^{2}-((1-\sigma_{2})\beta_{k})\|Y_{k}\|^{2}  +2nd(1+\beta_{0})\sigma_{2}^{2}\beta_{k}^{2}\sigma_{k}^{2}\cr
&+\frac{\beta_{0}+1}{(1-\sigma_{2})\beta_{k}}\alpha^{2}_{k}(8L^{2}\sum_{i=1}^{n}\|x_{i,k}-\overline{x}_{k}\|^{2}\cr
&+8nL^{2}\|\overline{x}_{k}-x^{*}\|^{2}+4\|\nabla f(X^{*})\|^{2}+\frac{2n\sigma_{g}^{2}}{\gamma_{k}}).
\end{align}

{\it Step 2:} From (\ref{CA3}) it follows that
\begin{align}\label{CA12}
&\|U_{k+1}\|^{2}
=\|\overline{x}_{k+1}-x^{*}\|^{2}\cr
=&\|(1-\beta_{k})\overline{x}_{k}-x^{*}+\beta_{k}(\overline{x}_{k}+\overline{n}_{k})\!-\!\frac{\alpha_{k}}{n}\sum_{i=1}^{n}g_{i}^{k}\|^{2}.
\end{align}
Recall that $\mathbb{E}[n_{k}|\mathcal{F}_{k}]=0$. Then, from (\ref{CA12}) it follows that
\begin{align}\label{CA12a}
&\mathbb{E}[\|U_{k+1}\|^{2}|\mathcal{F}_{k}]\cr
=&\mathbb{E}[\|\overline{x}_{k}-x^{*}-\frac{\alpha_{k}}{n}\sum_{i=1}^{n}g_{i}^{k}\|^{2}|\mathcal{F}_{k}]+\mathbb{E}[\|\beta_{k}\overline{n}_{k}\|^{2}|\mathcal{F}_{k}]\cr
=&\mathbb{E}[\|\overline{x}_{k}-x^{*}-\frac{\alpha_{k}}{n}\sum_{i=1}^{n}g_{i}^{k}+\frac{\alpha_{k}}{n}\sum_{i=1}^{n}\nabla f_{i}(x_{i,k})\cr
&-\frac{\alpha_{k}}{n}\sum_{i=1}^{n}\nabla f_{i}(x_{i,k})+\frac{\alpha_{k}}{n}\sum_{i=1}^{n}\nabla f_{i}(\overline{x}_{k})\cr
&-\frac{\alpha_{k}}{n}\sum_{i=1}^{n}\nabla f_{i}(\overline{x}_{k})\|^{2}|\mathcal{F}_{k}]+\mathbb{E}[\|\beta_{k}\overline{n}_{k}\|^{2}|\mathcal{F}_{k}].
\end{align}
From Assumption \ref{assum2}, we have
$\mathbb{E}[\|\frac{1}{n}\sum_{i=1}^{n}g_{i}^{k}-\frac{1}{n}\sum_{i=1}^{n}\nabla f_{i}(x_{i,k})\|^{2}|\mathcal{F}_{k}]\leq\frac{\sigma_{g}^{2}}{\gamma_{k}}.$
Then, from (\ref{CA12a}) it follows that
\begin{align}\label{CA13}
&\mathbb{E}[\|U_{k+1}\|^{2}|\mathcal{F}_{k}]\cr
\leq&\|\overline{x}_{k}-x^{*}-\frac{\alpha_{k}}{n}\sum_{i=1}^{n}\nabla f_{i}(\overline{x}_{k})\|^{2}\cr
&+\|-\frac{\alpha_{k}}{n}\sum_{i=1}^{n}\nabla f_{i}(x_{i,k})+\frac{\alpha_{k}}{n}\sum_{i=1}^{n}\nabla f_{i}(\overline{x}_{k})\|^{2}\cr
&+2\|\overline{x}_{k}-x^{*}-\frac{\alpha_{k}}{n}\sum_{i=1}^{n}\nabla f_{i}(\overline{x}_{k})\|\|-\frac{\alpha_{k}}{n}\sum_{i=1}^{n}\nabla f_{i}(x_{i,k})\cr
&+\frac{\alpha_{k}}{n}\sum_{i=1}^{n}\nabla f_{i}(\overline{x}_{k})\|+\mathbb{E}[\|\beta_{k}\overline{n}_{k}\|^{2}|\mathcal{F}_{k}]
+\frac{\alpha_{k}^{2}\sigma_{g}^{2}}{\gamma_{k}}.
\end{align}
Next, we analyze each term on the right-hand side of (\ref{CA13}).
\begin{align}\label{CA13a}
&\|-\frac{\alpha_{k}}{n}\sum_{i=1}^{n}\nabla f_{i}(x_{i,k})+\frac{\alpha_{k}}{n}\sum_{i=1}^{n}\nabla f_{i}(\overline{x}_{k})\|^{2}\cr
\leq&\frac{\alpha_{k}^{2}L^{2}}{n}\sum_{i=1}^{n}
\|x_{i,k}-\overline{x}_{k}\|^{2}.
\end{align}
Note that each function $f_{i}$ is $\mu$-strongly convex. Then, from Lemma 2.2 in \cite{Olshevsky2019} and there exists a sufficiently large $k_{0}>0$ such that
 $\alpha_{k}\leq\alpha_{k_{0}}\leq\frac{1}{L}$ for all $k>k_{0}$, it follows that
\begin{align}\label{CA13b}
\|\overline{x}_{k}-x^{*}-\frac{\alpha_{k}}{n}\sum_{i=1}^{n}\nabla f_{i}(\overline{x}_{k})\|^{2}
\leq(1-\mu\alpha_{k})^{2}\|\overline{x}_{k}-x^{*}\|^{2}.
\end{align}
Note that $\mathbb{E}[\|\overline{n}_{k}\|^{2}|\mathcal{F}_{k}]\leq2d\sigma_{k}^{2}$. Then, we have
\begin{eqnarray}\label{CA14}
\mathbb{E}[\|\beta_{k}\overline{n}_{k}\|^{2}|\mathcal{F}_{k}]
\leq2d\sigma_{k}^{2}\beta_{k}^{2}.
\end{eqnarray}
Thus, substituting (\ref{CA13a})-(\ref{CA14}) into (\ref{CA13}), we have
\begin{align}\label{CA20}
&\mathbb{E}[\|U_{k+1}\|^{2}|\mathcal{F}_{k}]\cr
\leq&(1-\mu\alpha_{k})^{2}\|\overline{x}_{k}-x^{*}\|^{2}+\frac{\alpha_{k}^{2}L^{2}}{n}\sum_{i=1}^{n}
\|x_{i,k}-\overline{x}_{k}\|^{2}\cr
&+2\frac{\alpha_{k}L(1-\mu\alpha_{k})}{\sqrt{n}}\|\overline{x}_{k}-x^{*}\|\sqrt{\sum_{i=1}^{n}\|x_{i,k}-\overline{x}_{k}\|^{2}}\cr
&+2d\sigma_{k}^{2}\beta_{k}^{2}+\frac{\alpha_{k}^{2}\sigma_{g}^{2}}{\gamma_{k}}\cr
=&\left((1-\mu\alpha_{k})\|\overline{x}_{k}-x^{*}\|+\frac{\alpha_{k}L}{\sqrt{n}}\sqrt{\sum_{i=1}^{n}
\|x_{i,k}-\overline{x}_{k}\|^{2}}\right)^{2}\cr
&+2d\sigma_{k}^{2}\beta_{k}^{2}+\frac{\alpha_{k}^{2}\sigma_{g}^{2}}{\gamma_{k}}.
\end{align}
By using Cauchy-Schwarz inequality with $\eta=\mu\alpha_{k}$. Note that there exists a sufficiently large $k_{1}>0$ such that
 $\alpha_{k}\leq\alpha_{k_{1}}\leq\frac{1}{\mu}$ for all $k>k_{1}$. Then, we have $(1+\eta)(1-\mu\alpha_{k})^{2}\leq(1-\mu\alpha_{k})$ and $(1+\frac{1}{\eta})\alpha_{k}\leq\frac{2}{\mu}$.
Thus, from (\ref{CA20}) it follows that
\begin{align}\label{CA21a}
&\mathbb{E}[\|U_{k+1}\|^{2}|\mathcal{F}_{k}]\cr
\leq&(1-\mu\alpha_{k})\|\overline{x}_{k}-x^{*}\|^{2}+\frac{2\alpha_{k}L^{2}}{n\mu}\sum_{i=1}^{n}
\|x_{i,k}-\overline{x}_{k}\|^{2}\cr
&+2d\sigma_{k}^{2}\beta_{k}^{2}+\frac{\alpha_{k}^{2}\sigma_{g}^{2}}{\gamma_{k}}.
\end{align}

{\it Step 3:} To establish the mean-square and almost-sure convergence of Algorithm 1, we introduce the following candidate of the Lyapunov function, which takes into account the time-scale difference between these two residual variables.
\begin{eqnarray}\label{CA24}
V(Y_{k},U_{k})=\|U_{k}\|^{2}+a_{k}\|Y_{k}\|^{2},
\end{eqnarray}
where $a_{k}=\frac{6L^{2}\alpha_{k}}{n(1-\sigma_{2})\mu\beta_{k}}$ is to characterize the time-scale difference between the two residual variables. For convenience, set $V_{k}=V(Y_{k},U_{k})$.

Note that $a_{k}$ is nonincreasing due to Assumption \ref{assum3}, namely, $a_{k+1}\leq a_{k}\leq a_{0}$. Then, from (\ref{CA10a}), (\ref{CA21a}) and (\ref{CA24}) it follows that
\begin{align}\label{CA25a}
&\mathbb{E}[V_{k+1}|\mathcal{F}_{k}]
=\mathbb{E}[\|U_{k+1}\|^{2}|\mathcal{F}_{k}]+a_{k+1}\mathbb{E}[\|Y_{k+1}\|^{2}|\mathcal{F}_{k}] \cr
\leq&\mathbb{E}[\|U_{k+1}\|^{2}|\mathcal{F}_{k}]+a_{k}\mathbb{E}[\|Y_{k+1}\|^{2}|\mathcal{F}_{k}] \cr
\leq&(1-\mu\alpha_{k}+\frac{8(a_{0}+1)(\beta_{0}+1)nL^{2}}{(1-\sigma_{2})}\frac{\alpha^{2}_{k}}{\beta_{k}})V_{k}\cr
&+((\mu\alpha_{k}-(1-\sigma_{2})\beta_{k})a_{k}+\frac{2\alpha_{k}L^{2}}{n\mu})\|Y_{k}\|^{2}\cr
&+2nd(1+\beta_{0})\sigma_{2}^{2}a_{k}\beta_{k}^{2}\sigma_{k}^{2}+2d\sigma_{k}^{2}\beta_{k}^{2}
+\frac{\alpha_{k}^{2}\sigma_{g}^{2}}{\gamma_{k}}\cr
&+\frac{\beta_{0}+1}{(1-\sigma_{2})}\frac{a_{k}\alpha^{2}_{k}}{\beta_{k}}(4\|\nabla f(X^{*})\|^{2}+\frac{2n\sigma_{g}^{2}}{\gamma_{k}}).
\end{align}
Note that $\sup_{k}\frac{\alpha_{k}}{\beta_{k}}\leq\min\{\frac{2(1-\sigma_{2})}{3\mu},\frac{(1-\sigma_{2})\mu}{16(a_{0}+1)(\beta_{0}+1)nL^{2}}\}$. Then, we have
\begin{align}\label{CA25b}
(\mu\alpha_{k}-(1-\sigma_{2})\beta_{k})a_{k}+\frac{2\alpha_{k}L^{2}}{n\mu}\leq0,\cr
-\mu\alpha_{k}+\frac{8(a_{0}+1)(\beta_{0}+1)nL^{2}}{(1-\sigma_{2})}\frac{\alpha^{2}_{k}}{\beta_{k}}\leq-\frac{\mu}{2}\alpha_{k}.
\end{align}
Further, from (\ref{CA25a}) and (\ref{CA25b}) it follows that
\begin{align}\label{CA25aa}
\mathbb{E}[V_{k+1}|\mathcal{F}_{k}]
\leq&V_{k}-\frac{\mu}{2}\alpha_{k}V_{k}\cr
&+2nd(1+\beta_{0})\sigma_{2}^{2}a_{k}\beta_{k}^{2}\sigma_{k}^{2}+2d\sigma_{k}^{2}\beta_{k}^{2}
+\frac{\alpha_{k}^{2}\sigma_{g}^{2}}{\gamma_{k}}\cr
&+\frac{\beta_{0}+1}{(1-\sigma_{2})}\frac{a_{k}\alpha^{2}_{k}}{\beta_{k}}(4\|\nabla f(X^{*})\|^{2}+\frac{2n\sigma_{g}^{2}}{\gamma_{k}}).
\end{align}
Therefore, by Assumption \ref{assum3} and Lemma \ref{lemCa2}, we have $V_{k}$ converges to 0 almost-surely, and $\sum_{k=0}^{\infty}\alpha_{k}V_{k}<\infty$, a.s.. The almost-sure convergence of the algorithm is obtained.

Taking expectations for both sides of (\ref{CA25aa}), we have
\begin{align}\label{CA25}
\mathbb{E}[V_{k+1}]
\leq&\mathbb{E}[V_{k}]-\frac{\mu}{2}\alpha_{k}\mathbb{E}[V_{k}]\cr
&+2nd(1+\beta_{0})\sigma_{2}^{2}a_{k}\beta_{k}^{2}\sigma_{k}^{2}+2d\sigma_{k}^{2}\beta_{k}^{2}
+\frac{\alpha_{k}^{2}\sigma_{g}^{2}}{\gamma_{k}}\cr
&+\frac{\beta_{0}+1}{(1-\sigma_{2})}\frac{a_{k}\alpha^{2}_{k}}{\beta_{k}}(4\|\nabla f(X^{*})\|^{2}+\frac{2n\sigma_{g}^{2}}{\gamma_{k}}).
\end{align}
Therefore, by Assumption \ref{assum3} and Lemma \ref{lemCa2}, we have $\mathbb{E}[V_{k}]$ converges to 0 almost-surely, and $\sum_{k=0}^{\infty}\alpha_{k}\mathbb{E}[V_{k}]<\infty$, a.s.. The mean-square convergence of the algorithm is also obtained. $\hfill\Box$

Next, we show how the added privacy noise affects the convergence rate of the algorithm.
\begin{theorem}\label{them7}
If Assumptions \ref{assum1}-\ref{assum2a} hold, and  $\alpha_{k}=\frac{a_{1}}{(k+a_{2})^{\alpha}}$, $\beta_{k}=\frac{a_{1}}{(k+a_{2})^{\beta}}, $
$\gamma_{k}=\lceil a_{3}(k+a_{2})^{\gamma}\rceil$ and $\sigma_{k}=O((k+a_{2})^{\eta})$, $a_{1}, a_{2}, a_{3}>0$, $0<\beta<\alpha\leq1$, $0\leq\gamma$, $0\leq\eta\leq\frac{3\beta-2}{2}$, then the convergence rate of Algorithm 1 is given as follows:
When $0<\alpha<1$, there holds $\mathbb{E}[\|x_{i,k}-x^{*}\|^{2}]=O\big(\frac{1}{(k+a_{2})^{\min\{3\beta-2\alpha-2\eta,\alpha-\beta\}}}\big).$
When $\alpha=1$, there holds $\mathbb{E}\big[\|x_{i,k}-x^{\ast}\|^{2}\big]=O\big(\frac{\ln k}{(k+a_{2})^{\min\{a_{1}\mu-1+\beta,3\beta-2\eta-2,1-\beta \}}}\big),$
where $\mu$ is a positive constant in Assumption \ref{assum1}.
\end{theorem}
\textit{Proof}: Set $\alpha_{k}=\frac{a_{1}}{(k+a_{2})^{\alpha}}$, $\beta_{k}=\frac{a_{1}}{(k+a_{2})^{\beta}}$, $\gamma_{k}=\lceil a_{3}(k+a_{2})^{\gamma}\rceil$, and $\sigma_{k}=O((k+a_{2})^{\eta})$, $0<\beta<\alpha\leq1$, $0\leq\gamma$, $0\leq\eta\leq\frac{3\beta-2}{2}$. Then, for large enough $k_{0}$, there exist constants $C_{0}>0$, $C_{1}>0$, $C_{2}>0$, and $C_{3}>0$, from (\ref{CA25}) it follows that
\begin{align}\label{thm71}
 &\mathbb{E}\big[V_{k+1}\big]\cr
\leq&\big[1-\frac{a_{1}\mu}{(k+a_{2})^{\alpha}}+\frac{C_{0}}{(k+a_{2})^{2\alpha-\beta}}\big]\mathbb{E}\big[V_{k_{0}}\big]+\frac{C_{1}}{(k+a_{2})^{2\alpha}}\cr
&+\frac{C_{2}}{(k+a_{2})^{2\beta-2\eta}}
+\frac{C_{3}}{(k+a_{2})^{3\alpha-2\beta}}, ~~{\rm as}~~k>k_{0}.
\end{align}
Note that $0\leq\eta\leq\frac{3\beta-2}{2}<\beta$ and $\beta<\alpha$. Then, $2\beta-2\eta<2\alpha$, and from (\ref{thm71}) it follows that
\begin{align*}
 &\mathbb{E}\big[V_{k+1}\big]\cr
\leq&\big[1-\frac{a_{1}\mu}{(k+a_{2})^{\alpha}}+\frac{C_{0}}{(k+a_{2})^{2\alpha-\beta}}\big]\mathbb{E}\big[V_{k_{0}}\big]\cr
&+\frac{C_{2}}{(k+a_{2})^{2\beta-2\eta}}
+\frac{C_{3}}{(k+a_{2})^{3\alpha-2\beta}}, ~~{\rm as}~~k>k_{0}.
\end{align*}
Thus, by iterating the above process, we have
\begin{align}\label{CA30}
&\mathbb{E}\big[V_{k+1}\big]\cr
\leq&\prod_{t=k_{0}}^{k}\big[1-\frac{a_{1}\mu}{(t+a_{2})^{\alpha}}+\frac{C_{0}}{(t+a_{2})^{2\alpha-\beta}}\big]\mathbb{E}\big[V_{k_{0}}\big]\cr
&+\sum_{l=k_{0}}^{k-1}\prod_{t=l+1}^{k}(1-\frac{a_{1}\mu}{(t+a_{2})^{\alpha}}+\frac{C_{0}}{(t+a_{2})^{2\alpha-\beta}})\frac{C_{2}}{l^{2\beta-2\eta}}\cr
&+\sum_{l=k_{0}}^{k-1}\prod_{t=l+1}^{k}(1-\frac{a_{1}\mu}{(t+a_{2})^{\alpha}}+\frac{C_{0}}{(t+a_{2})^{2\alpha-\beta}})\frac{C_{3}}{l^{3\alpha-2\beta}}\cr
&+\frac{C_{2}}{(k+a_{2})^{2\beta-2\eta}}+\frac{C_{3}}{(k+a_{2})^{3\alpha-2\beta}}.
\end{align}
When $\alpha=1$, since $2-\beta>1$, from (\ref{est:prod}) it follows that
$\prod_{t=l+1}^{k}(1-\frac{a_{1}\mu}{t+a_{2}}+\frac{C_{0}}{(t+a_{2})^{2-\beta}})
=O \left(\left( \frac{l+a_{2}}{k+a_{2}} \right)^{a_{1}\mu}\right).$
Further, from Lemma \ref{lemma:prod_est2}, we have
\begin{align*}
&\sum_{l=k_{0}}^{k-1}\prod_{t=l+1}^{k}(1-\frac{a_{1}\mu}{t+a_{2}}+\frac{C_{0}}{(t+a_{2})^{2-\beta}})\frac{C_{2}}{l^{2\beta-2\eta}}\cr
=&\left\{
\begin{array}{ccc}
O(\frac{1}{(k+a_{2})^{a_{1}\mu}}),  && 2\beta-2\eta-1>a_{1}\mu;  \\
O(\frac{\ln k}{(k+a_{2})^{a_{1}\mu}}),  && 2\beta-2\eta-1=a_{1}\mu;   \\
O(\frac{1}{(k+a_{2})^{2\beta-2\eta-1}}), && 2\beta-2\eta-1<a_{1}\mu;   \\
\end{array}
\right.\cr
&\sum_{l=k_{0}}^{k-1}\prod_{t=l+1}^{k}(1-\frac{a_{1}\mu}{t +a_{2}}+\frac{C_{0}}{(t+a_{2})^{2-\beta}})\frac{C_{3}}{l^{3-2\beta}}\cr
=&\left\{
\begin{array}{ccc}
O(\frac{1}{(k+a_{2})^{a_{1}\mu}}),  && 2-2\beta>a_{1}\mu;  \\
O(\frac{\ln k}{(k+a_{2})^{a_{1}\mu}}),  && 2-2\beta =a_{1}\mu;   \\
O(\frac{1}{(k+a_{2})^{2-2\beta}}), && 2-2\beta<a_{1}\mu.   \\
\end{array}
\right.
\end{align*}
This further implies that
\begin{align}\label{CA30a}
\mathbb{E}\big[V_{k+1}\big]=O\big(\frac{\ln k}{(k+a_{2})^{\min\{a_{1}\mu,2\beta-2\eta-1,2-2\beta\}}}\big).
\end{align}
Note the selection of $V_{k}$. Then, this together with (\ref{CA30a}) implies the result.

When $0<\alpha<1$, note that $\beta<\alpha$ and $k_0$ is large enough. Then, for any $k>k_0$, we have
$-\frac{a_{1}\mu}{(k+a_{2})^{\alpha}}+\frac{C_{0}}{(k+a_{2})^{2\alpha-\beta}}
\leq -\frac{a_{1}\mu}{2(k+a_{2})^{\alpha}}.$
From (\ref{ineq:prod}) it follows that
\begin{align*}
&\prod_{t=k_{0}}^{k}\big[1-\frac{a_{1}\mu}{2(t+a_{2})^{\alpha}}\big]\cr
=&O\big(\exp\big(-\sum_{t=k_{0}}^{k}\frac{a_{1}\mu}{2(t+a_{2})^{\alpha}}\big)\big)\cr
=&O\big(\exp\big(-\frac{a_{1}\mu}{2(1-\alpha)}[(k+a_{2}+1)^{1-\alpha}-(k_0+a_{2})^{1-\alpha}]\big)\big).
\end{align*}
Note that for large enough $k_0$ and $l\geq k_0$, we have $(1-\frac{a_{1}\mu}{2(l+a_{2})^{\alpha}})^{-1}\leq 2$. Therefore, we have
\begin{align*}
&\sum_{l=k_{0}}^{k-1}\prod_{t=l+1}^{k}(1-\frac{a_{1}\mu}{(t+a_{2})^{\alpha}}+\frac{C_{0}}{(k+a_{2})^{2\alpha-\beta}})\frac{C_{2}}{l^{2\beta-2\eta}}\cr
\leq&\sum_{l=k_{0}}^{k-1}\prod_{t=l+1}^{k}(1-\frac{a_{1}\mu}{2(t+a_{2})^{\alpha}})\frac{C_{1}}{(l+a_{2})^{2\beta-2\eta}}\cr
\leq&2\sum_{l=k_{0}}^{k-1}\prod_{t=l}^{k}(1-\frac{a_{1}\mu}{2(t+a_{2})^{\alpha}})\frac{C_{1}}{(l+a_{2})^{2\beta-2\eta}}\cr
=&O\big(\sum_{l=k_{0}}^{k-1}\exp\big(-\frac{a_{1}\mu}{2(1-\alpha)}(k+a_{2}+1)^{1-\alpha}\big)\cr
&\qquad\cdot\exp\big(\frac{a_{1}\mu}{2(1-\alpha)}(l+a_{2})^{1-\alpha}\big)\frac{C_{1}}{(l+a_{2})^{2\beta-2\eta}}\big).
\end{align*}
From Lemma \ref{lemma:calc_sum_exp} and (\ref{CA30}), we further have
\begin{align*}
&\mathbb{E}\big[V_{k+1}\big]\cr
=&O\big(\exp\big(-\frac{a_{1}\mu}{2(1-\alpha)}(k+a_{2}+1)^{1-\alpha}\big)\big)\cr
&+O\big(\exp\big(-\frac{a_{1}\mu}{2(1-\alpha)}(k+a_{2}+1)^{1-\alpha}\big)\cr
&\qquad\cdot \frac{1}{(k+a_{2})^{2\beta-\alpha-2\eta}}\exp\big(\frac{a_{1}\mu}{2(1-\alpha)}(k+a_{2})^{1-\alpha}\big) \big) \cr
& +O\big(\exp\big(-\frac{a_{1}\mu}{2(1-\alpha)}(k+a_{2}+1)^{1-\alpha}\big)\cr
&\qquad\cdot \frac{1}{(k+a_{2})^{2\alpha-2\beta}}\exp\big(\frac{a_{1}\mu}{2(1-\alpha)}(k+a_{2})^{1-\alpha}\big) \big) \cr
&+O\big(\frac{1}{(k+a_{2})^{2\beta-2\eta}}\big)+O\big(\frac{1}{(k+a_{2})^{3\alpha-2\beta}}\big)\cr
=& O\big(\frac{1}{(k+a_{2})^{2\beta-\alpha-2\eta}}\big)+O\big(\frac{1}{(k+a_{2})^{2\alpha-2\beta}}\big).
\end{align*}
This together with the definition of $V_{k}$ implies the result. $\hfill\Box$
\begin{remark}
Inspired by the linear two-time-scale stochastic approximation in \cite{Doan2021c}, the almost-sure and mean-square convergence of the algorithm with $\sigma_{k}=O((k+a_{2})^{\eta})$, $0\leq\eta\leq\frac{3\beta-2}{2}$, is studied by properly choosing a Lyapunov function. Based on this, the convergence rate of the algorithm is given in Theorem \ref{them7}, and the related results are not provided for distributed stochastic optimization even when no privacy protection is considered. Note that the convergence rate for distributed optimization with non-vanishing noises is studied in \cite{Doan2021a,Reisizadeh2022}, where $\sigma_{k}=O((k+a_{2})^{\eta})$, $\eta=0$. Then, the convergence rate studied in this paper is nontrivial and more general than the one in \cite{Doan2021a,Reisizadeh2022}.
\end{remark}

From Theorems \ref{thm2} and \ref{them7}, the mean-square convergence of Algorithm 1 and differential privacy with a finite cumulative privacy budget $\varepsilon$ for an infinite number of iterations can be simultaneously established, which will be shown in the following corollary:
\begin{corollary}\label{corollary1} Let $\alpha_{k}=\frac{a_{1}}{(k+a_{2})^{\alpha}}$, $\beta_{k}=\frac{a_{1}}{(k+a_{2})^{\beta}}$, $\gamma_{k}=\lceil a_{3}(k+a_{2})^{\gamma}\rceil$, and $\sigma_{k}=\underline{b}(k+a_{2})^{\eta}$, $0<a_1<a_2^\beta$, $a_2, a_{3},\underline{b}>0$. If $\alpha+\gamma-\beta+\eta>1$, $0<\beta<\alpha\leq1$, $0\leq\gamma$, $0\leq\eta\leq\frac{3\beta-2}{2}$ hold, then the mean-square convergence of Algorithm 1 and differential privacy with a finite cumulative privacy budget $\varepsilon$ for an infinite number of iterations are established simultaneously.
\end{corollary}
\begin{remark} Corollary \ref{corollary1} holds when the added privacy noises have an increasing variance. For example, when $\alpha=1$, $\beta=0.9$, $\gamma=1.06$, $\eta=0.35$, or $\alpha=0.9$, $\beta=0.8$, $\gamma=1.8$, $\eta=0.2$, the conditions of Corollary \ref{corollary1} hold. In this case, the mean-square convergence of Algorithm 1 and differential privacy with a finite cumulative privacy budget $\varepsilon$ for an infinite number of iterations can be established simultaneously. Note that $\varepsilon$-differential privacy is proven only for one iteration, leading to a cumulative privacy loss of $k\varepsilon$ after $k$ iterations \cite{Wang2022d,Li2018,Ding2021b}. Then, Algorithm 1 is superior to the ones in \cite{Wang2022d,Li2018,Ding2021b}.
\end{remark}
\begin{remark}
Our approach does not contradict the trade-off between utility and privacy in the differential-privacy theory. In fact, to achieve differential privacy, our approach does incur a cost (compromise) on the utility. However, different from existing approaches which compromise convergence accuracy to enable differential privacy, our approach compromises the convergence rate (which is also a utility metric) instead. From Theorem \ref{them7} it follows that  the convergence rate of the algorithm slows down with the increase of the privacy noise parameters. The ability to retain convergence accuracy makes our approach suitable for accuracy-critical scenarios.
\end{remark}
\section{Differentially private distributed stochastic optimization via gradient-perturbation}
This section presents a gradient perturbation method for privacy-preserving distributed stochastic optimization algorithms with time-varying sample sizes, i.e., Algorithm 2. Different from Algorithm 1, each agent $i$ updates its state as follows:
$x_{i,k+1}=(1-\beta_{k})x_{i,k}+\beta_{k}\sum_{j\in\mathcal{N}_{i}}a_{ij}x _{j,k}-\alpha_{k}(g_{i}^{k}+n_{i,k}),$
where $n_{i,k}\in\mathbb{R}^{d}$ is the added privacy noises for Agent $i$ at each time $k$, and is temporally and spatially independent.
\subsection{Privacy analysis}
In Algorithm 2, the privacy noise $n_{i,k}$ is added directly to the gradient. Then, the sensitivity of Algorithm 2 is $\Delta_{k}=\frac{1}{\gamma_{k}}\|g_{i}(x_{i,k},\xi_{i}^{l_{0}})-g_{i}(x_{i,k}, \xi_{i}^{l'_{0}})\|_{1}\leq \frac{C}{\gamma_{k}}.$
\begin{theorem} Let $C$ be any given positive number. If
 $\varepsilon=\sum_{k=1}^{\infty}\frac{C}{\gamma_{k}\sigma_{k}},$
then Algorithm 2 is $\varepsilon$-differentially private for an infinite number of iterations. Furthermore, if $\sigma_{k}=O((k+a_{2})^{\eta})$, $\gamma_{k}=\lceil a_{3}(k+a_{2})^{\gamma}\rceil$ with $\eta+\gamma>1$, $a_{2}, a_{3}>0$, then Algorithm 2 is differentially private with a finite cumulative privacy budget $\varepsilon$ for an infinite number of iterations.
\end{theorem}
\noindent
\textit{Proof}: The results can be obtained similar to the proof process of Theorem \ref{thm1} with $\Delta_{k}\leq\frac{C}{\gamma_{k}}$, and differential privacy is robust to post-processing as shown in Proposition 2.1 of \cite{Dwork2006}. $\hfill\Box$
\subsection{Convergence analysis}
For convergence analysis, we need the following assumptions about the step sizes $\alpha_{k}, \beta_{k}$, privacy noise parameters $\sigma_{k}$, and time-varying sample sizes $\gamma_{k}$.
\begin{assumption}\label{assum5} The step sizes $\alpha_{k}, \beta_{k}$, privacy noise parameters $\sigma_{k}$, and time-varying sample sizes $\gamma_{k}$ satisfy the following conditions:
\begin{itemize}
\item
$\sup_{k}\frac{\alpha_{k}}{\beta_{k}}\!\!\leq\!\!\min\{\frac{2(1-\sigma_{2})}{3\mu},\frac{(1-\sigma_{2})^{2}\mu^{2}\beta_{0}}{16(6L^{2}\alpha_{0}+n(1-\sigma_{2})\mu\beta_{0})(\beta_{0}+1)L^{2}}\},\\
\sum_{k=0}^{\infty}\frac{\alpha^{2}_{k}}{\beta_{k}}<\infty,
\sum_{k=0}^{\infty}\frac{\alpha^{2}_{k}}{\gamma_{k}\beta_{k}}<\infty,
\sum_{k=0}^{\infty}\frac{\alpha^{2}_{k}\sigma_{k}^{2}}{\beta_{k}}<\infty,\\
\sum_{k=0}^{\infty}\frac{\alpha^{2}_{k}}{\gamma_{k}}<\infty,
\sum_{k=0}^{\infty}\alpha_{k}^{2}\sigma_{k}^{2}<\infty.$
\end{itemize}
\end{assumption}
\begin{remark} For example, for sufficiently large $a_{2}$, Assumption \ref{assum5} is satisfied in the form of $\alpha_{k}=(k+a_{2})^{-1}$, $\beta_{k}=(k+a_{2})^{-\beta}$, $\beta\in(1/2,1)$, $\sigma_{k}=(k+a_{2})^{\eta}, \eta<(1-\beta)/2$, $\gamma_{k}=\lceil(k+a_{2})^{\gamma}\rceil, \gamma\geq0$.
\end{remark}

Next, we provide the mean-square and almost sure convergence of Algorithm 2.
\begin{theorem}
If Assumptions \ref{assum1}-\ref{assum2a} and \ref{assum5} hold, then Algorithm 2 converges in mean-square and almost surely for any $i\in \mathcal{V}$.
\end{theorem}
\noindent
\textit{Proof}: The proof is similar to that of Theorem \ref{alth3}. And thus, here we only present the main different parts as follows:

{\it Step 1:} (\ref{CA10a}) is replaced by
\begin{align*}
\mathbb{E}[\|Y_{k+1}\|^{2}|\mathcal{F}_{k}]
\leq&(1-(1-\sigma_{2})\beta_{k})\|Y_{k}\|^{2}+\frac{(\beta_{0}+1)}{(1-\sigma_{2})\beta_{k}}\alpha^{2}_{k}\cr
&(8L^{2}\sum_{i=1}^{n}\|x_{i,k}-\overline{x}_{k}\|^{2}
+8nL^{2}\|\overline{x}_{k}-x^{*}\|^{2}\cr
&+4\|\nabla f(X^{*})\|^{2}+\frac{2n\sigma_{g}^{2}}{\gamma_{k}}+2nd\sigma_{k}^{2}).
\end{align*}

{\it Step 2:} (\ref{CA21a}) is replaced by
$\mathbb{E}[\|U_{k+1}\|^{2}|\mathcal{F}_{k}] \leq(1-\mu\alpha_{k})\|U_{k}\|^{2}+\frac{2\alpha_{k}L^{2}}{n\mu}\|Y_{k}\|^{2}+\frac{\alpha_{k}^{2}\sigma_{g}^{2}}{\gamma_{k}}+2d^{2}\alpha_{k}^{2}\sigma_{k}^{2}.
$.

{\it Step 3:} (\ref{CA25}) is replaced by
\begin{align*}
&\mathbb{E}[V_{k+1}|\mathcal{F}_{k}] \cr
\leq&(1-\mu\alpha_{k}+\frac{8(a_{0}+1)(\beta_{0}+1)nL^{2}}{(1-\sigma_{2})}\frac{\alpha^{2}_{k}}{\beta_{k}})V_{k}\cr
&+2d\sigma_{k}^{2}\alpha_{k}^{2}+\frac{\alpha_{k}^{2}\sigma_{g}^{2}}{\gamma_{k}}
+\frac{\beta_{0}+1}{(1-\sigma_{2})}\frac{a_{k}\alpha^{2}_{k}}{\beta_{k}}(4\|\nabla f(X^{*})\|^{2}\cr
&+\frac{2n\sigma_{g}^{2}}{\gamma_{k}}+2nd\sigma_{k}^{2}).
\end{align*}
Similar to that of Theorem \ref{alth3}, by Assumption \ref{assum5} and Lemma \ref{lemCa2}, the result is obtained. $\hfill\Box$
\begin{theorem} If Assumptions~\ref{assum1}-\ref{assum2a} hold, and $\alpha_{k}=\frac{a_{1}}{(k+a_{2})^{\alpha}}$, $\beta_{k}=\frac{a_{1}}{(k+a_{2})^{\beta}},$
$\gamma_{k}=\lceil a_{3}(k+a_{2})^{\gamma}\rceil$ and $\sigma_{k}=O((k+a_{2})^{\eta})$, $a_{1}, a_{2}, a_{3}>0$, $0<\beta<\alpha\leq1$, $0\leq\gamma$, $0\leq\eta\leq \min\{\frac{\beta}{2},\frac{\alpha-\beta}{2}\}$, then the convergence rate of Algorithm 2 is given as follows:
When $0<\alpha<1$, there holds
$\mathbb{E}[\|x_{i,k}-x^{*}\|^{2}]=O\big(\frac{1}{(k+a_{2})^{\min\{\beta-2\eta,\alpha-\beta-2\eta\}}}\big).$
When $\alpha=1$, there holds
$\mathbb{E}\big[\|x_{i,k}-x^{\ast}\|^{2}\big]=O\big(\frac{\ln k}{(k+a_{2})^{\min\{a_{1}\mu-1+\beta,\beta-2\eta,1-\beta-2\eta\}}}\big),$
where $\mu$ is a positive constant in Assumption \ref{assum1}.
\end{theorem}
\noindent
\textit{Proof}: We replace $2\beta-2\eta$ with $2\alpha-2\eta$, and $3\alpha-2\beta$ with $3\alpha-2\beta-2\eta$ in Theorem \ref{them7}, and the result can be obtained similar to the proof of Theorem \ref{them7}. $\hfill\Box$

\begin{corollary}\label{corollary2} Let $\alpha_{k}=\frac{a_{1}}{(k+a_{2})^{\alpha}}$, $\beta_{k}=\frac{a_{1}}{(k+a_{2})^{\beta}}$, $\gamma_{k}=\lceil a_3(k+a_{2})^{\gamma}\rceil$, and $\sigma_{k}=\underline{b}(k+a_{2})^{\eta}$, $a_1, a_2, a_3,\underline{b}>0$. If $\gamma+\eta>1$, $0<\beta<\alpha\leq1$, $0\leq\gamma$, $0\leq\eta\leq \min\{\frac{\beta}{2}, \frac{\alpha-\beta}{2}\}$ hold, then the mean-square convergence of Algorithm 2 and differential privacy with a finite cumulative privacy budget $\varepsilon$ for an infinite number of iterations are established simultaneously.
\end{corollary}
\begin{remark} For example, when we choose $\alpha=1$, $\beta=0.6$, $\gamma=1.1$, and $\eta=0.1$, Corollary \ref{corollary2} holds. Note that the mean square convergence of the proposed algorithm cannot be guaranteed \cite{Xu2021,Ding2021b}. Then, Algorithm 2 is superior to the ones in \cite{Xu2021,Ding2021b}.
\end{remark}
\subsection{Oracle complexity analysis}
Based on Theorem 7, we establish the oracle (sample) complexity  for obtaining an $\epsilon$-optimal solution satisfying $\mathbb{E}[\|x_{i,k}-x^{*}\|^{2}]\leq\epsilon$, where $\epsilon>0$ is sufficiently small. The oracle complexity, measured by the total number of sampled gradients for deriving an $\epsilon$-optimal solution, is $\sum_{k=0}^{K(\epsilon)}\gamma_{k}$, where $K(\epsilon)=\min_{k}\{k:\mathbb{E}[\|x_{i,k}-x^{*}\|^{2}]\leq\epsilon\}$.
 \begin{corollary}\label{corollary3}
 If Assumptions \ref{assum1}-\ref{assum2a} hold, and  $\alpha_{k}=\frac{a_{1}}{(k+1)^{\alpha}}$, $\beta_{k}=\frac{a_{1}}{(k+1)^{\beta}}, $
$\gamma_{k}=\lceil a_{3}(k+1)^{\gamma}\rceil$ and $\sigma_{k}=\underline{b}(k+1)^{\eta}$, $0<a_{1}<1$, $a_{3},\underline{b}>0$, $\beta=0.5+\epsilon,$ $\alpha=1-\epsilon$, $\gamma=\epsilon$, $\eta=\epsilon$, then the oracle complexity  of Algorithm 2 is $O\big(\epsilon^{-\frac{1+\epsilon}{0.5-4\epsilon}}\big)$.
\end{corollary}
\noindent
\textit{Proof}: Similar to the proof of Theorem 7, there exists a constant $C_{1}$ such that $\mathbb{E}[\|x_{i,k}-x^{*}\|^{2}]\leq C_{1} k^{-(0.5-4\epsilon)}$, and hence,  $K(\epsilon)=(\frac{C_{1}}{\epsilon})^{\frac{1}{0.5-4\epsilon}}$. Thus, the oracle complexity is $\sum_{k=0}^{K(\epsilon)}\gamma_{k}=\sum_{k=0}^{K(\epsilon)}\lceil a_{3}(k+1)^{\gamma}\rceil=O\big(\epsilon^{-\frac{1+\epsilon}{0.5-4\epsilon}}\big)$.  $\hfill\Box$
\begin{remark}
The increasing sample size schemes can generally be employed only when sampling is relatively cheap compared to the communication burden \cite{Lei2020} or the main computational step, such as  computing a projection or a prox \cite{Cui2023}. As $k$ becomes large, one might question how one deals with $\gamma_{k}$ tending to $+\infty$. This issue does not arise in machine learning due to $\epsilon$-optimal solution is interested; e.g. if $\epsilon=10^{-3}$, then such a scheme requires approximately $O(10^{6})$ samples in total from Corollary \ref{corollary3}. Such requirements are not terribly onerous particularly since the computational cost of centralized stochastic gradient descent is $O(10^{6})$ to achieve the same accuracy as our scheme. In addition, for the finite sample space, when the samples required by this scheme are larger than the total samples, the convergence can be guaranteed by setting the required samples equal to the total samples.
\end{remark}
\section{Example}
This section shows the efficiency and advantages of Algorithms 1-2 on distributed parameter estimation problems and distributed training of a convolutional neural network over ``MNIST" datasets.

In distributed parameter estimation problems, we consider a network of $n$ spatially distributed sensors that aim to estimate an unknown $d$-dimensional parameter $x^{*}$. Each sensor $i$ collects a set of scalar measurements $d_{i,l}$ generated by the following linear regression model corrupted with noises, $d_{i,l}=u_{i,l}^{T} x^{*}+n_{i,l},$ where $u_{i,l}\in\mathbb{R}^{d}$ is the regression vector accessible to Agent $i$, and $n_{i,l}\in\mathbb{R}$ is a zero-mean Gaussian noise. Suppose that $u_{i,l}$ and $n_{i,l}$ are mutually independent Gaussian sequences with distributions $N\left(\mathbf{0}, R_{u, i}\right)$ and $N\left(0, \sigma_{i, \nu}^{2}\right)$, respectively. Then, the distributed parameter estimation problem can be modeled as a distributed stochastic quadratic optimization problem, $\min \sum_{i=1}^{n} f_{i}(x),$ where $f_{i}(x)=\mathbb{E}\left[\left\|d_{i,l}-u_{i,l}^{T} x\right\|^{2}\right].$ Thus, $f_{i}(x)=\left(x-x^{*}\right)^{T} R_{u, i}\left(x-x^{*}\right)+\sigma_{i,v}^{2}$ is convex and $\nabla f_{i}(x)=R_{u, i}(x-x^{*})$. By using the observed regressor $u_{i,l}$ and the corresponding measurement $d_{i,l}$, the sampled gradient $u_{i,l} u_{i,l}^{T} x-d_{i,l} u_{i,l}$ satisfies Assumption \ref{assum2}. Set the vector dimension $d=6$ and the true parameter $x^{*}=\frac{{\bf 1}}{2}$. Let $n=6$; the adjacency matrix of the communication graph satisfies Assumption 3. In addition, the initial parameter estimates of these agents are chosen as $x_{i,0}=[3,1,1,3,3,1]^{T}$, $i=1,2,3,4,5,6$. Let each covariance matrix $R_{u, i}=\left[
            \begin{array}{cccccc}
              2 & 1 & 0 & 1 & 0 &0\\
              1 & 2 & 0 & 1 &0 &0\\
              0 & 0 & 2 & 0 &0 &0\\
              1 & 1 & 0 & 2 &0 &0 \\
              0 & 0 & 0 & 0 &2 &0\\
              0 & 0 & 0 & 0 &0 &2
            \end{array}
          \right]$
be positive definite. Then, each $f_{i}(x)$ is strong convex.
\begin{figure}
\begin{center}
\subfigure[Algorithm 1]{\includegraphics[width=0.24\textwidth]{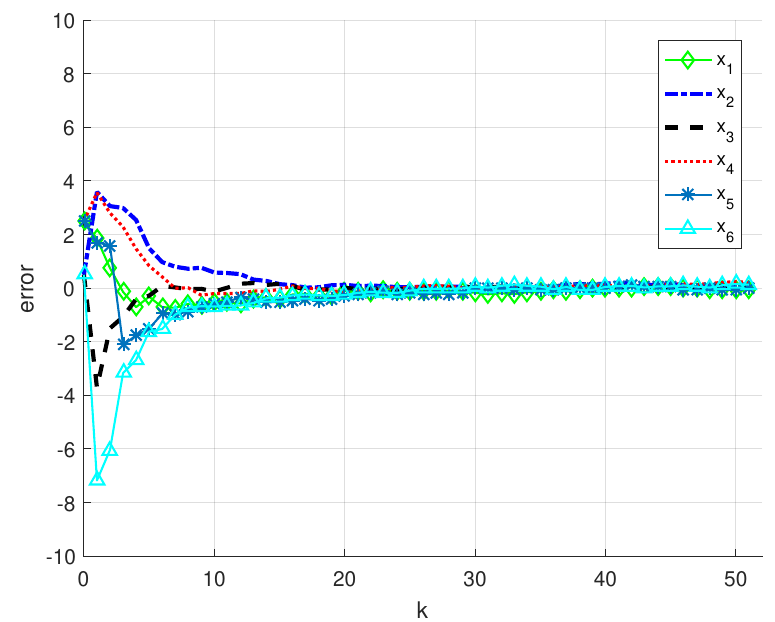}}
\subfigure[Algorithm 2]{\includegraphics[width=0.24\textwidth]{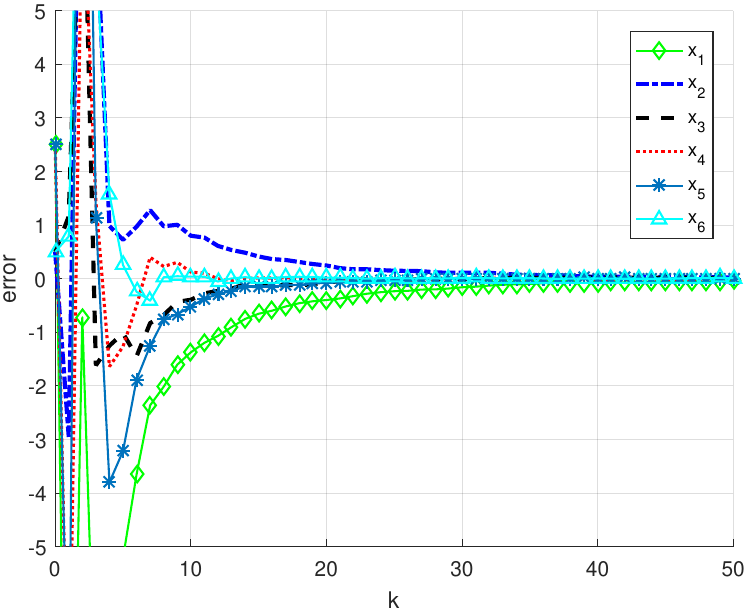}}
\caption{Convergence}
\end{center}
\label{fig2}
\end{figure}
\begin{figure}
\begin{center}
\subfigure[Algorithm 1]{\includegraphics[width=0.24\textwidth]{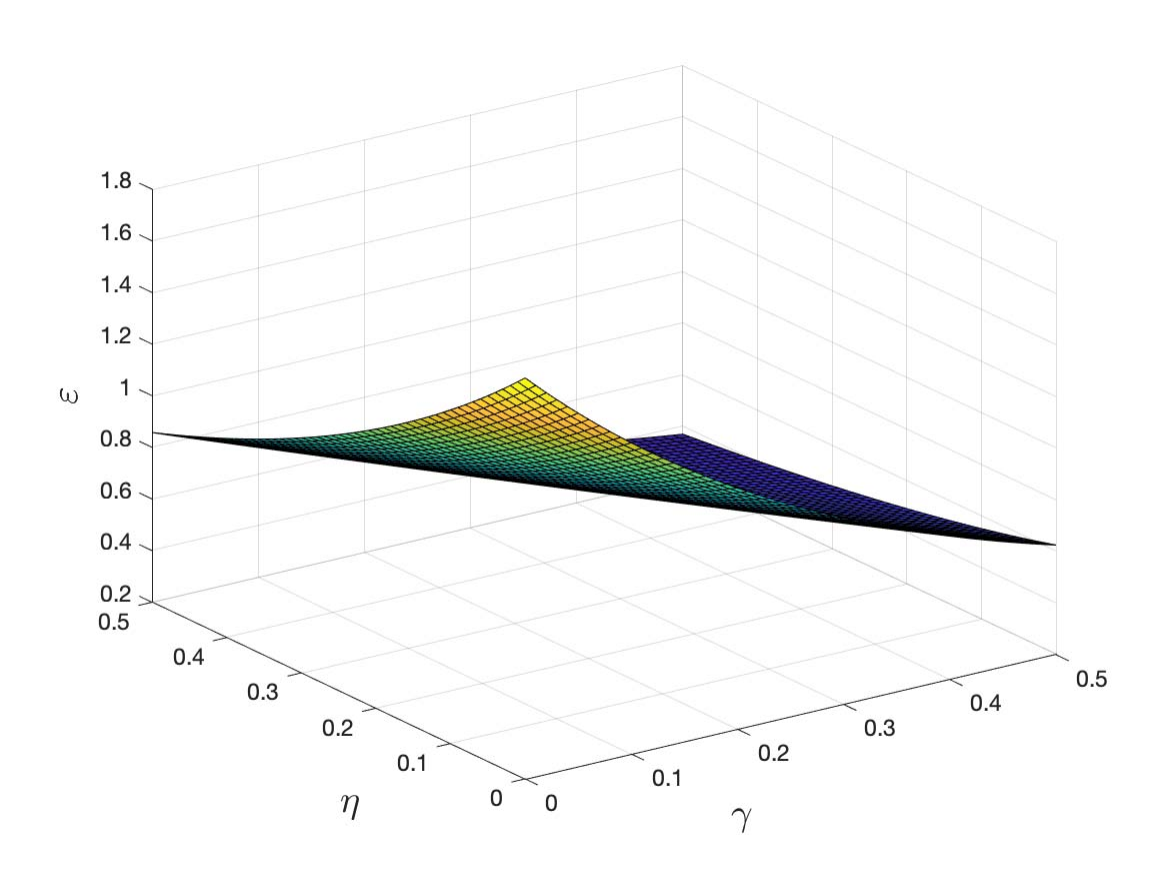}}
\subfigure[Algorithm 2]{\includegraphics[width=0.24\textwidth]{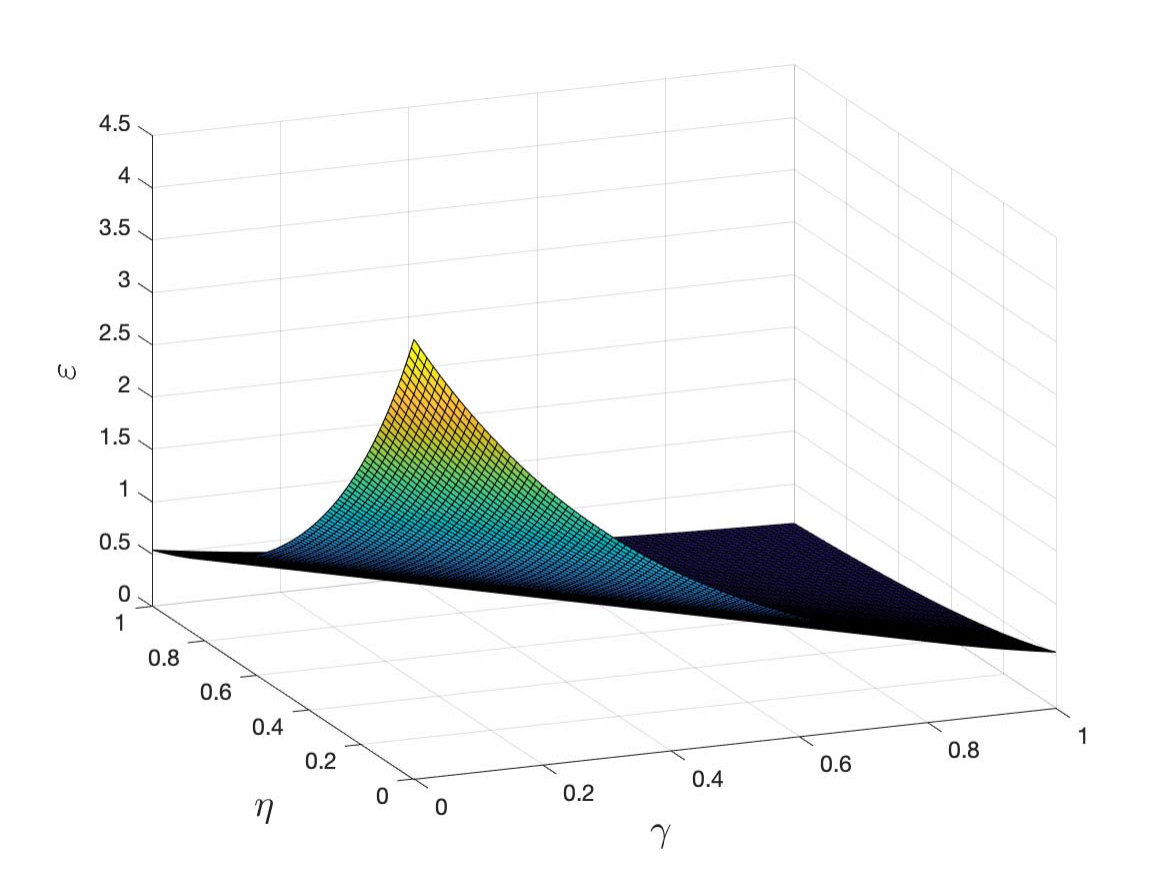}}
\caption{The relationship between $\varepsilon$, $\eta$, and $\gamma$}
\end{center}
\end{figure}
First, we set $C=0.2$, the step size $\alpha_{k}=0.5/(k+1)^{0.9}$, $\beta_{k}=0.5/(k+1)^{0.6}$, the sample size $\gamma_{k}=\lceil(k+1)^{1.1}\rceil$, and the privacy noise parameter $\sigma_{k}=(k+1)^{0.05}$. Then, the cumulative privacy budget for an infinite number of iterations is finite with $\varepsilon\approx0.864$. The estimation error of Algorithm 1 is displayed in Fig. 1 (a), showing that the generated iterations asymptotically converge to the true parameter $x^{*}$. Second, we set $C=0.2$, the step size ${\alpha_k}=0.5/(k+1)^{0.8}$ and $\beta_{k}=0.5/(k+1)^{0.5}$, the sample size $\gamma_{k}=\lceil(k+1)^{1.2}\rceil$, and the privacy noise parameter $\sigma_{k}=(k+1)^{0.1}$. Then, the cumulative privacy budget for an infinite number of iterations is finite with $\varepsilon\approx0.488$. The estimation error of Algorithm 2 is illustrated in Fig. 1 (b), showing that the generated iterations asymptotically converge to the true parameter $x^{*}$.

For both algorithms, we show the situation that $\varepsilon$ is affected by $\eta$ and $\gamma$ in Fig. 2. As shown, $\varepsilon$ decreases with the increase of $\eta$ and $\gamma$, which is consistent with the theoretical analysis.

{\bf Comparison with the existing works:} The comparison between Algorithm 1 and  [42], [44] is shown in Fig. 3; the comparison between Algorithm 2 and [43], [44] is shown in Fig. 4, respectively. From Fig. 3, the mean-square convergence of Algorithm 1 and differential privacy with a finite cumulative privacy budget $\varepsilon$ for an infinite number of iterations are established simultaneously, but the algorithm in [42], [44] cannot achieve the above results. From Fig. 4, the mean-square convergence of Algorithm 2 and differential privacy with a finite cumulative privacy budget $\varepsilon$ for an infinite number of iterations are established simultaneously, but the algorithm in [43], [44] cannot achieve the above results. Based on the above discussions, Algorithms 1-2 achieve higher accuracy while keeping high-level privacy protection compared to [42], [43], [44].

{\bf Distributed training on a benchmark machine learning dataset:} We evaluate the performance of Algorithm 1 through distributed training of a convolutional neural network (CNN) using the ``MNIST" dataset. Specifically, 5 agents collaboratively train a CNN model on a communication graph, and the adjacency matrix satisfies Assumption 3. The ``MNIST" dataset is uniformly divided into 5 pieces, each of which is sent to an agent. At each iteration, a time-varying batch of samples is drawn from each agent's local dataset by the bootstrapping method. The CNN model has 2 convolutional layers, and each layer has 16 and 32 filters, respectively, followed by a max pooling layer. The Sigmoid function is used as the activation function, and hence Assumption 1 is satisfied. Then, the output is flattened and sent to a fully connected layer for 10 classes. We set the noise parameters ${{\sigma }_{k}}={{(k+2)}^{0.01}}$, step-sizes ${{\alpha }_{k}}=\frac{0.01}{{{(k+2)}^{0.76}}}$, ${{\beta }_{k}}=\frac{0.01}{{{(k+2)}^{0.51}}}$, and time-varying sample sizes ${{\gamma }_{k}}=\lceil{{(k+2)}^{1.4}}\rceil$. The validation accuracy of 5 agents after 2000 iterations is given in Fig. 5 (a). Then, the comparison of Algorithm 1 and [42] is given in Fig. 5 (b). To ensure the initial conditions, the same noise parameters and communication graph are used with the step-sizes $\alpha =0.01$ and the batch size $B=50$. From Fig. 5 (b), it can be seen that the validation accuracy of Algorithm 1 is over $80\%$ after 2000 iterations, but [42] cannot train the CNN model well.
\begin{figure}
\begin{center}
\subfigure[Convergence accuracy]{\includegraphics[width=0.24\textwidth]{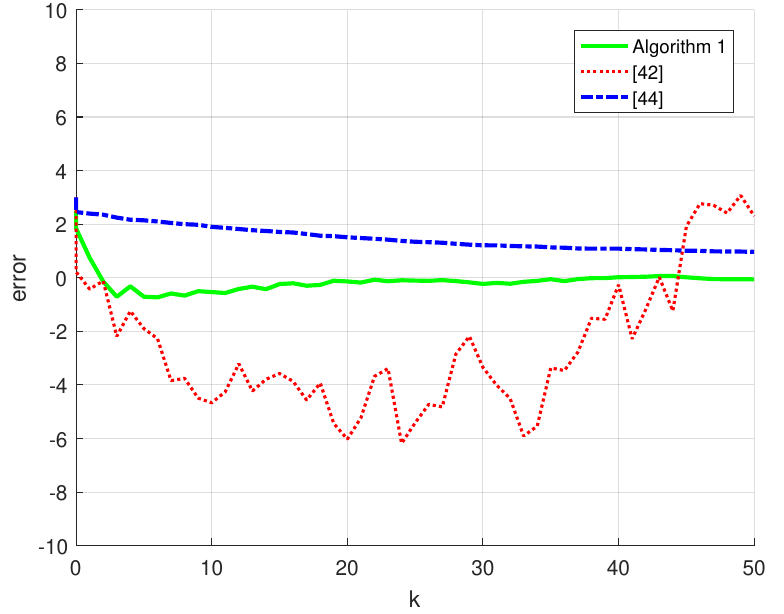}}
\subfigure[Privacy level]{\includegraphics[width=0.24\textwidth]{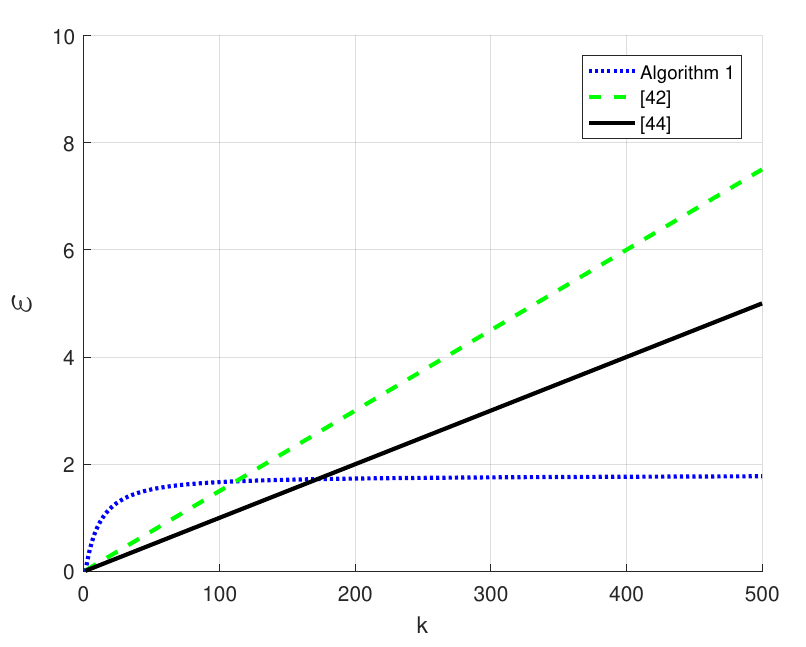}}
\caption{Comparison between Algorithm 1 and the existing works}
\end{center}
\end{figure}
\begin{figure}
\begin{center}
\subfigure[Convergence accuracy]{\includegraphics[width=0.24\textwidth]{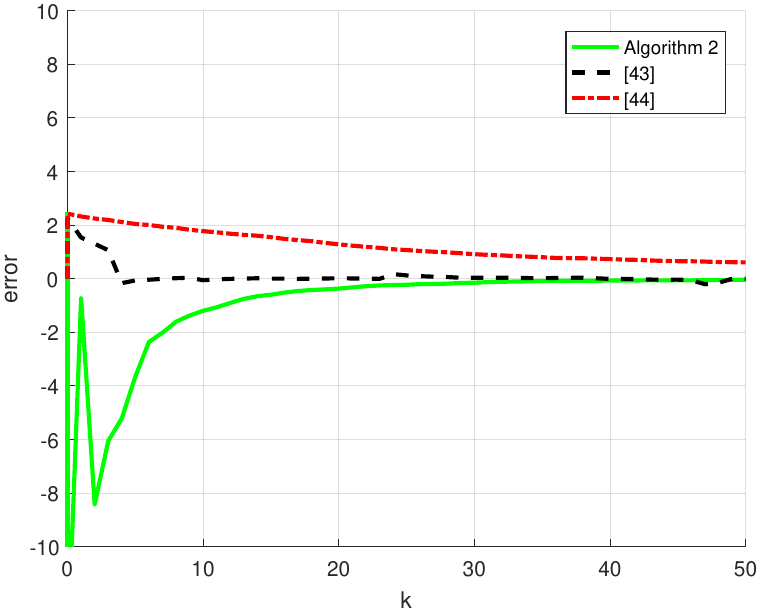}}
\subfigure[Privacy level]{\includegraphics[width=0.24\textwidth]{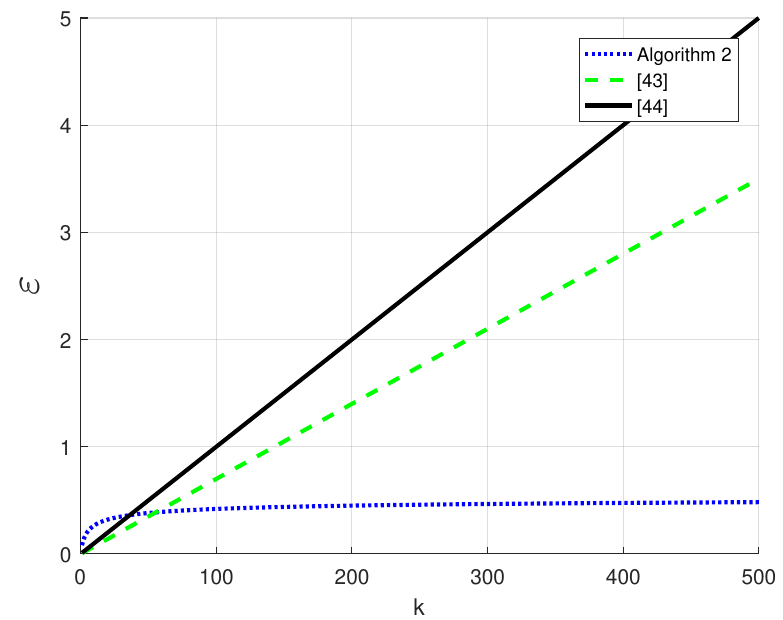}}
\caption{Comparison between Algorithm 2 and the existing works}
\end{center}
\end{figure}
\begin{figure}
\begin{center}
\subfigure[Training accuracy]{\includegraphics[width=0.24\textwidth]{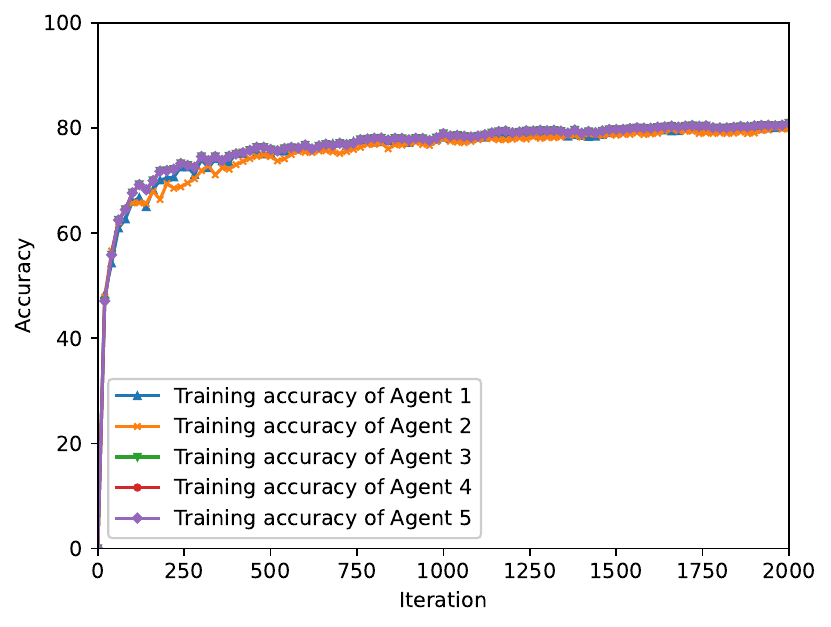}}
\subfigure[Comparison]{\includegraphics[width=0.24\textwidth]{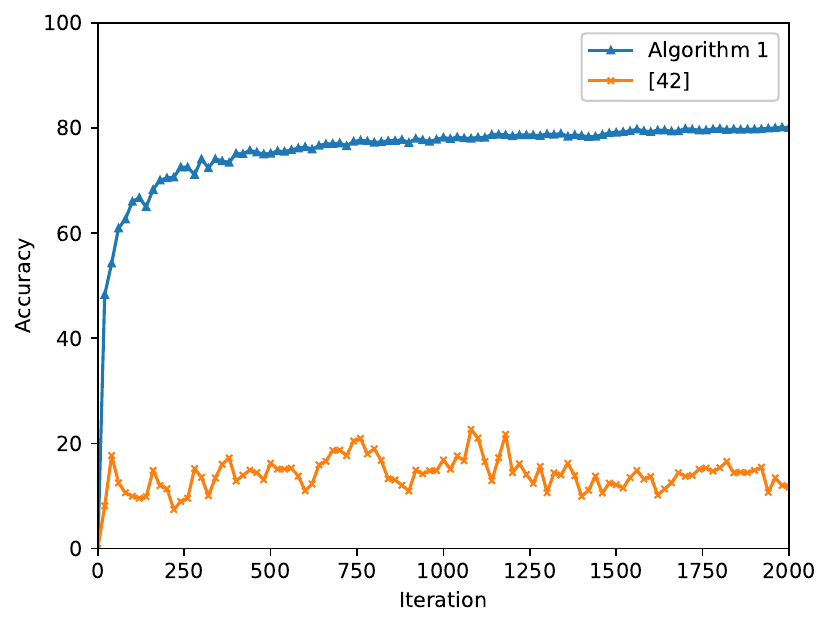}}
\caption{Training accuracy of Algorithm 1 using the ``MNIST" dataset}
\end{center}
\label{fig5}
\end{figure}

\section{Concluding remarks}
Two differentially private distributed stochastic optimization algorithms with time-varying sample sizes have been studied in this paper. Both gradient- and output-perturbation methods are employed. By using two-time scale stochastic approximation-type conditions, the algorithm converges to the optimal point in an almost sure and mean-square sense and is simultaneously differentially private with a finite cumulative privacy budget $\varepsilon$ for an infinite number of iterations. Furthermore, it is shown how the added privacy noise affects the convergence rate of the algorithm. Finally, numerical examples including distributed training over ``MNIST" datasets are provided to verify the efficiency of the algorithms. In the future, we will consider the privacy-preserving of other distributed stochastic optimization algorithms, including distributed alternating direction method of multipliers, distributed gradient tracking methods and distributed stochastic dual averaging.
\section*{Appendix A. Lemmas}\label{appendix-0}

\setcounter{equation}{0}
\setcounter{lemma}{0}
\renewcommand{\theequation}{A.\arabic{equation}}
\newtheorem{lemmaxx}{Lemma}
\renewcommand{\thelemma}{A.\arabic{lemma}}
\begin{lemma}\cite{WangJM2022}\label{lemma:calc_sum_exp} For any given $c, \ k_0\geq0,\ 0<p\leq1,$ and $q\in \mathbb{R}$, we have
$
\sum_{l=1}^{k} \frac{\exp\left(c(l+k_0)^p\right)}{(l+k_0)^q} =  O\left( \frac{\exp\left( c(k+k_0)^p \right)}{(k+k_0)^{p+q-1}} \right) .
$
\end{lemma}
\begin{lemma}\label{lemma:prod_est} For $ 0<\beta\leq 1,\ \alpha>0,\ k_0\geq0$, sufficiently large $l$, we have
\begin{align}\label{ineq:prod}
&\prod_{i=l}^k \left(1-\frac{\alpha}{(i+k_0)^\beta}\right)\nonumber\\
\leq& \!\!
\begin{cases}
\left( \frac{l+k_0}{k+k_0} \right)^\alpha, & \beta =1;\\
\exp\left( \frac{\alpha}{1-\beta}\left((l\!+\!k_0)^{1-\beta}\!-\!(k\!+\!k_0\!+\!1)^{1-\beta}\right) \right), & \!\!\!\beta\in(0,1).
		\end{cases}
	\end{align}
If we further assume that $\rho>0$, then for any $ \gamma>0 $, we have
	\begin{align}\label{est:prod}
\prod_{i=l}^k \left(1-\frac{\alpha}{i+k_0}+\frac{\gamma}{(i+k_0)^{1+\rho}}\right)
=O \left(\left( \frac{l+k_0}{k+k_0} \right)^\alpha\right).
	\end{align}
\end{lemma}
\textit{Proof}: \eqref{ineq:prod} is obtained from Lemma 1.2 in \cite{WangJM2022}.

Note that
\begin{align}\label{eq:prod*prod}
	&\prod_{i=l}^k \left(1-\frac{\alpha}{i+k_0}+\frac{\gamma}{(i+k_0)^{1+\rho}}\right)\nonumber\\
	=& \prod_{i=l}^k \left(1-\frac{\alpha}{i+k_0}\right) \prod_{i=l}^k \left(1+O\left(\frac{1}{(i+k_0)^{1+\rho}}\right)\right).
\end{align}
Since $\rho>0$, by Theorem 2.1.3 of \cite{Pan2015Order}, we have
$
\sup_{l,k}\prod_{i=l}^k \left(1+O\left(\frac{1}{(i+k_0)^{1+\rho}}\right)\right)<\infty,
$
which together with \eqref{ineq:prod} and \eqref{eq:prod*prod} implies \eqref{est:prod}.  $\hfill\Box$

\begin{lemma}\cite{Ke2023}\label{lemma:prod_est2} For the sequence $h_{k}$, assume that (i) $h_{k}$ is positive and monotonically increasing; (ii) $\ln h_{k} = o(\ln k)$. Then, for real numbers $a_{1}$, $a_{2}$, $\chi$, and any positive integer $p$,
\begin{align*}
\sum_{l=1}^{k}\prod_{i=l+1}^{k}\left(1-\frac{a_{1}}{i+a_{2}}\right)^{p}\frac{h_{l}}{l^{1+\chi}}
=\left\{
\begin{array}{lcc}
O\left(\frac{1}{k^{pa_{1}}}\right),  & pa_{1}<\chi;  \\
 O\left(\frac{h_{k}\ln k}{k^{\chi}}\right),  & pa_{1}=\chi;   \\
 O\left(\frac{h_{k}}{k^{\chi}}\right), & pa_{1}>\chi.   \\
  \end{array}
  \right.
\end{align*}
\end{lemma}
\begin{lemma} \cite{Goodwin1984}. \label{lemCa2} Let $V_{k}$, $u_{k}$, $\beta_{k}$, $\zeta_{k}$ be non-negative random variables. If $\sum_{k=0}^{\infty}u_{k}<\infty$,$\sum_{k=0}^{\infty}\beta_{k}<\infty$, and $\mathbb{E}[V_{k+1}|\mathcal{F}_{k}]\leq(1+u_{k})V_{k}-\zeta_{k}+\beta_{k}$ for all $k\geq0$, then $V_{k}$ converges almost surely and $\sum_{k=0}^{\infty}\zeta_{k}<\infty$ almost surely. Here $\mathbb{E}[V_{k+1}|\mathcal{F}_{k}]$ denotes the conditional mathematical expectation for the given $V_{0},\ldots,V_{k}$, $u_{0},\ldots,u_{k}$, $\beta_{0},\ldots,\beta_{k}$, $\zeta_{0},\ldots,\zeta_{k}$.
\end{lemma}
\begin{lemma}\label{lemCa3}
For a matrix $A\in\mathbb{R}^{n\times n}$ with eigenvalues $\lambda_{1}\geq\cdots\geq\lambda_{n}$ and corresponding non-zero mutually orthogonal eigenvectors $v_{1},\ldots, v_{n}$. If a vector $u\in\mathbb{R}^{n}$ is orthogonal to $v_{1},\ldots, v_{m-1}$ for some $m\leq n$, then $\|Au\|\leq\lambda_{m}\|u\|$.
\end{lemma}
\noindent
\textit{Proof}: Under the condition of the lemma, vector $u$ can be written as $u=\alpha_{m}v_{m}+\cdots+\alpha_{n}v_{n}$. Therefore, one can get
\begin{eqnarray*}
\frac{\|Au\|}{\|u\|}=\sqrt{\frac{\alpha_{m}^{2}\|v_{m}\|^{2}\lambda_{m}^{2}+\cdots+\alpha_{n}^{2}\|v_{n}\|^{2}\lambda_{n}^{2}}{\alpha_{m}^{2}\|v_{m}\|^{2}+\cdots+\alpha_{n}^{2}\|v_{n}\|^{2}}}\leq \lambda_{m},
\end{eqnarray*}
which implies the lemma. $\hfill\Box$

\end{document}